\documentclass{article}

\usepackage{PRIMEarxiv}
\usepackage[numbers]{natbib}
\usepackage[utf8]{inputenc} 
\usepackage[T1]{fontenc}    
\usepackage{hyperref}       
\usepackage{url}            
\usepackage{booktabs}       
\usepackage{amsfonts}       
\usepackage{nicefrac}       
\usepackage{microtype}      
\usepackage{lipsum}
\usepackage{fancyhdr}       
\usepackage{graphicx}       
\usepackage{newtxtext}
\usepackage{multicol}
\usepackage{multirow}
\usepackage{subcaption}
\usepackage{color}
\usepackage{amsthm}
\usepackage{newtxmath}
\usepackage{soul}
\usepackage{hyperref}
\usepackage{titling}
\usepackage{comment}
\usepackage{amsthm,amsmath}

\usepackage{mathtools}
\usepackage{etoolbox}

\DeclareFontFamily{U}{matha}{\hyphenchar\font45}
\DeclareFontShape{U}{matha}{m}{n}{
<-6> matha5 <6-7> matha6 <7-8> matha7
<8-9> matha8 <9-10> matha9
<10-12> matha10 <12-> matha12
}{}
\DeclareSymbolFont{matha}{U}{matha}{m}{n}

\DeclareFontFamily{U}{mathx}{\hyphenchar\font45}
\DeclareFontShape{U}{mathx}{m}{n}{
<-6> mathx5 <6-7> mathx6 <7-8> mathx7
<8-9> mathx8 <9-10> mathx9
<10-12> mathx10 <12-> mathx12
}{}
\DeclareSymbolFont{mathx}{U}{mathx}{m}{n}

\DeclareMathDelimiter{\vvvert} {0}{matha}{"7E}{mathx}{"17}%

\DeclarePairedDelimiterX{\normiii}[1]
{\vvvert}
{\vvvert}
{\ifblank{#1}{\:\cdot\:}{#1}}

\newcommand{\tX}{\tilde{X}}

\newcommand{\ty}{\tilde{y}}

\newcommand{\tbeta}{\tilde{\beta}}

\newcommand{\hatDelta}{\hat{\Delta}}

\newcommand{\tr}{\mathrm{tr}}

\newcommand{\diag}{\mathrm{diag}}

\newcommand{\Sp}{\mathcal{S}}

\newtheorem{lemma}{Lemma}[section]

\newtheorem{corollary}{Corollary}[section]
\newtheorem{proposition}{Proposition}[section]
\newtheorem{assumption}{Assumption}[section]

\makeatother

\title{\huge{$\ell_1$-Regularized Generalized Least Squares}}
\author{Kaveh S. Nobari$^{1,2,\dagger}$ \& Alex Gibberd$^{3,*,\dagger}$ \vspace{0.3cm} \\
    \small{$^1$} \small{Centre for Climate Finance and Investment, Imperial College London, UK}\\
    \small{$^2$} \small{Department of Psychological and Behavioural Science, London
School of Economics, UK}
    \\ 
    $^3$ \small{School of Mathematical Sciences, Lancaster University, UK} \vspace{0.1cm} \\
\footnotesize{$^*$ Correspondence to: a.gibberd@lancaster.ac.uk} \vspace{0.1cm}\\
\footnotesize{$^\dagger$ Equal contribution}}
\date{}

\begin{document}

\maketitle

\begin{abstract}
We study an $\ell_{1}$-regularized generalized least-squares (GLS)
estimator for high-dimensional regressions with autocorrelated errors.
Specifically, we consider the case where errors are assumed to follow
an autoregressive process, alongside a feasible variant of GLS that
estimates the structure of this process in a data-driven manner. The
estimation procedure consists of three steps: performing a LASSO regression,
fitting an autoregressive model to the realized residuals, and then
running a second-stage LASSO regression on the rotated (whitened)
data. We examine the theoretical performance of the method in a sub-Gaussian
random-design setting, in particular assessing the impact of the rotation
on the design matrix and how this impacts the estimation error of
the procedure. We show that our proposed estimators maintain smaller
estimation error than an unadjusted LASSO regression when the errors
are driven by an autoregressive process. A simulation study verifies
the performance of the proposed method, demonstrating that the penalized
(feasible) GLS-LASSO estimator performs on par with the LASSO in the
case of white noise errors, whilst outperforming when the errors exhibit
significant autocorrelation.
\end{abstract}

\keywords{High-Dimensional, Regression, Generalized Least Squares, Autocorrelation, Regularization}

\section{Introduction \label{Introduction}}

When performing regression in the high-dimensional setting, where
the number of covariates $p$ is greater than the number of data-points
$n$, it is common to utilize regularized estimators that constrain
parameters to lie in some restricted sub-space. Consider the linear
regression model 
\begin{equation}
y=X\beta+Lu\;,\label{eq: regression}
\end{equation}
where $X$ is a high-dimensional matrix of stochastic explanatory
variables, and $L$ is a lower-triangular matrix such that $e=Lu$
represents a sequence of random errors with autocovariance matrix
$\mathrm{\Gamma:=Cov}(e)=\sigma_{u}^{2}LL^{\top}$. If we assume the
coefficients $\beta$ are sparse, i.e. many are zero, then a popular
estimator for these coefficients is given by the \emph{least-absolute
shrinkage and selection operator (LASSO hereafter) }defined as

\begin{align}
\hat{\beta} & =\arg\min_{\beta\in\mathbb{R}^{p}}\left[\frac{1}{2n}\|y-X\beta\|_{2}^{2}+\lambda_{n}\|\beta\|_{1}\right],\label{eq:lasso}
\end{align}
for $\lambda_{n}\geq0.$ There has been much research on such estimators
(see \cite{wainwright2019high,buhlmann2011statistics} for a review),
including many alternative forms of regularization, e.g. ridge regression
(Tikhonov regularization), elastic-net, and the group-LASSO penalty,
among others. Broadly speaking, different regularization functions
enable one to easily impose restrictions or priors on the parameters
\cite{hastie2009elements,Negahban2012,Wainwright2009}. The focus
of this paper is to improve such estimators in the presence of autocorrelated
noise where we focus on the $\ell_{1}$ penalized case as an example.

If one considers (\ref{eq:lasso}), then we would intuitively expect
the estimator to be more efficient when faced with data generated
with independent errors, compared to when the errors possess some
general autocorrelation structure\footnote{For a given stationary variance $\diag(\Gamma)=\sigma^{2}$.}.
Generalized Least Squares (GLS) represents a simple extension to the
least-squares objective that can mitigate this increased variation,
whereby we \emph{whiten} the data prior to performing the estimation,
i.e., we define $\tilde{y}=Ry$ and $\tilde{X}=RX$ where $R=L^{-1}$---thus
after whitening, the errors $\tilde{y}-\tilde{X}\beta$ will appear
to have zero correlation. In a classical $p$ fixed, $n\rightarrow\infty$
asymptotic setting, GLS estimators have been well studied theoretically
\cite{Kariya2004,Koreisha2001}. For our study, we consider the GLS-LASSO
estimator of the following form:
\begin{equation}
\tilde{\beta}=\arg\min_{\beta\in\mathbb{R}^{p}}\left[\frac{1}{2n}\|R(y-X\beta)\|_{2}^{2}+\tilde{\lambda}_{n}\|\beta\|_{1}\right]\;,\label{eq:regGLS}
\end{equation}
where $\tilde{\lambda}_{n}$ is distinct from $\lambda_{n}$ in (\ref{eq:lasso}).
In practice, we need to estimate the covariance structure of the errors,
resulting in a \emph{feasible }GLS-LASSO (FGLS) where we replace $R$
with $\hat{R}$. In the classical setting, if we can provide a consistent
estimator of the whitening matrix $R$, then the GLS attains the best
linear unbiased estimator (BLUE) status and satisfies the Cramer-Rao
lower bound \cite{white2014asymptotic}.

The perils of autocorrelated errors within a least-squares framework
and spurious regression are famously discussed in \cite{box1971some}
and \cite{granger1974spurious}. Whilst there has been considerable
research that considers applications of the LASSO to time-series,
to our knowledge little work has been done to investigate how to correct
for autocorrelation, to enable more efficient finite sample estimation.
Examples of work in the general time-series setting include the study
of asymptotic robustness to autocorrelated errors \cite{Kock2014},
and applications of the LASSO to vector auto-regressive models \cite{Basu2015}.
There has also been work looking at how heteroskedasticity can be
taken into account when using the LASSO, e.g., when $\Gamma$ is diagonal,
but has time-varying entries \cite{Wagener2013,Ziel2016,Medeiros2016,Jia2015}.
In contrast to these studies, we here look at how one can correct
for autocorrelated errors and thus potentially improve the efficiency
of our estimator. As a motivating application, one could consider
regressing say, asset returns against some financial indicator, such
as the dividend-to-price ratio. In such circumstances, the returns
may often be considered stationary, whereas the dividend ratio may
be integrated of order $d\in\mathbb{N}$, or be a long-memory and
fractionally integrated process with $1/2\leq d<1$, and infinite
variance \cite{granger1974spurious,granger1981some}. Whilst we do
not study this case exactly, we do consider the case where the error
process of the regression may be highly persistent, i.e. the errors
are stationary but have a very large variance. To investigate this
scenario, we let $e_{t}$ be an auto-regressive AR($q$) process and
allow its parameters $\phi\in\mathbb{R}^{q}$ to be close to the boundary
of the stationary region. For example, in the AR(1) setting, we could
let the AR parameter $|\phi|$ be close to unity and hence study situations
where $\mathrm{Var}[e_{t}]\propto(1-\phi^{2})^{-1}$ will be large.
Our interest will be on how well the GLS procedure can adjust for
the impact of autocorrelation in the estimation of sparse regression
coefficients. 

The advantage of working with this simple error assumption is that
we can carefully track the errors incurred at each stage of the regularized
GLS procedure. Specifically, one of our contributions is to understand
how the error incurred by the first-stage LASSO estimate impacts the
subsequent (feasible) estimation of the autocorrelation structure,
via the error incurred in the estimation of the AR parameters. We
then look at how the estimated whitening matrix $\hat{R}$ impacts
the performance of the second-stage estimator (\ref{eq:regGLS}) for
which we provide a bound on the estimation error. Our main result
shows that in the autoregressive error situation the feasible GLS-LASSO
is able to provide consistent and efficient estimation. In particular,
as long as the sample size is sufficiently large, the GLS-LASSO (and
FGLS variant) can recover performance similar to that of a LASSO regression
faced with white-noise errors. This contrasts with the results of
\cite{Jia2015} who show poor performance of the GLS-LASSO with certain
kinds of heteroskadisticity, and underlines the importance of mapping
how the matrix $\hat{R}$ impacts the eigenvalues of the design matrix.
We pay particular attention to the so-called \emph{restricted eigenvalue}
(RE) conditions required for the LASSO, and demonstrate that even
though the whitening matrix induces dependence in the rows of $\hat{R}X$,
the RE condition can still be maintained in high-probability. We demonstrate
this in the setting where both the errors and the design matrix are
assumed sub-Gaussian, extending the work of \cite{Rudelson2012,Yu2010}
to settings where the design matrix can be correlated both cross-sectionally
and across samples. 

Readers should note, that as we were working on this paper, we came
across \cite{chronopoulos2023high} who propose the same GLS procedure
as here. Although the estimators are the same, our paper has a slightly
different focus. Importantly, rather than imposing the assumption
that a restricted eigenvalue on $\hat{R}X$ holds, we study precisely
the impact of the rotation matrix $\hat{R}$ on $X$ and how this
modifies the random-design matrix, with a particular interest of showing
when the RE conditions will still be satisfied. We take particular
care to study the impact of $\mathrm{Var}[e_{t}]$ in any constants
throughout our arguments, resulting in error bounds that are typically
predicated on maintaining sufficient samples $n=\Omega(\mathrm{Var}[e_{t}]\log p)$. 

The structure of the paper is as follows: Section \ref{Framework and notations}
provides the theoretical framework of the LASSO estimator in terms
of error bounds and its empirical behavior in the presence of autocorrelated
errors. Section \ref{sec:rgls} introduces the GLS extension of the
LASSO estimator, we present a result on the eigenvalues of the GLS
design matrix in a sub-Gaussian setting, and subsequently provide
an oracle inequality for the GLS-LASSO in the Gaussian design setting.
Bounding the error of the first-stage LASSO, we derive a bound on
the error incurred by $\hat{\phi}$, based on the estimated residuals.
We use this bound to derive conditions on the strength of the second-stage
regularizer, and thus provide an error bound for the FGLS-LASSO procedure.
Finally, Section \ref{sec:Experimental-Results} assesses the empirical
performance of the LASSO, GLS-LASSO and the FGLS-LASSO estimators
in terms of estimation error and sign recovery, before providing some
conclusions and directions for future work in Section \ref{sec:Conclusion}.

\subsection*{Notation}

Let $[p]:=\{1,\cdots,p\}$, and the support of the vector $x\in\mathbb{R}^{p}$
to be $\mathrm{supp}(x)=\{i\in[p]\:\vert\:x_{i}\ne0\}$. The $\ell_{q}$
norm of a vector is denoted $\|x\|_{q}=(\sum_{i=1}^{p}\lvert x_{i}\rvert^{q})^{1/q}$.
Matrix norms are denoted as $\|X\|_{F}:=(\sum_{ij}\lvert X_{ij}\rvert^{2})^{1/2}$,
$\|X\|_{\infty}:=\max_{ij}\lvert X_{ij}\rvert$, $\|X\|_{2}:=\sup_{\|v\|_{2}\le1}\|Xv\|_{2}$,
and $|X|_{\infty}:=\max_{j}\|A_{j\cdot}\|_{1}$. Letting $\Sp\subseteq[p]$,
we refer to $x_{\Sp}$ as the vector formed by this subset of elements,
with the rest of the elements set to zero, i.e. $x_{\Sp}=(x_{i}\;\mathrm{if}\;i\in\Sp,x_{i}=0\;\mathrm{otherwise})_{i=1}^{p}$.
We define the sub-Gaussian and sub-exponential norms of the random
variable $z$ according to $\normiii{z}_{2}=\inf\{t>0\:|\;\mathbb{E}[e^{z^{2}/t^{2}}]\le2\}$,
and $\normiii{z}_{1}=\inf\{t>0\:|\;\mathbb{E}[e^{|z|/t}]\le2\}$.
We write $f(n)=\mathcal{O}(g(n))$ if $f(n)\le c_{1}g(n)$ and $f(n)=\Omega(g(n))$
if $f(n)\ge c_{2}g(n)$ for some finite $c_{1},c_{2}>0$.

\section{The LASSO with Dependent Errors \label{Framework and notations}}

In this section, we provide a discussion of how the LASSO behaves
when faced with autocorrelated (autoregressive) errors. As one might
expect, we can see in certain situations that when the autocorrelation
function of the errors decays slowly in the case of persistent errors,
then the LASSO will experience a significant increase in estimation
error. To motivate the GLS correction we give theoretical and empirical
evidence for this increase in error. In the next section, we review
the estimation error incurred by the LASSO\footnote{For those interested in further details, the book of \cite{wainwright2019high}
and paper of \cite{Negahban2012} are highly recommended for analysis
of M-estimators, with the LASSO as a key example.}, importantly, we study the projection of the error process onto the
covariates, and show how the scale of this projection motivates a
choice of $\lambda_{n}$ and ultimately leads to error bounds for
the LASSO.

\subsection{Error Bounds for the LASSO}

Given the LASSO (\ref{eq:lasso}) is an M-estimator, the estimator
$\hat{\beta}$ necessarily results in a lower objective over the sample
data than the true parameters, such that $\mathrm{Loss}(\hat{\beta};\{x_{t},y_{t}\}_{t=1}^{n})\le\mathrm{Loss}(\beta;\{x_{t},y_{t}\}_{t=1}^{n})$.
Letting $\hat{\Delta}:=\hat{\beta}-\beta$, after some algebra one
can show

\begin{equation}
0\leq\frac{1}{n}\lVert X\hatDelta\rVert_{2}^{2}\leq\frac{2}{n}(X\hatDelta)^{\top}Lu+2\lambda_{n}\{\lVert\beta\rVert_{1}-\lVert\hat{\beta}\rVert_{1}\}\;.\label{eq: rev3}
\end{equation}
Now, consider that the $\ell_{1}$ norm is linearly decomposable over
distinct indices, i.e. we have $\|\beta\|_{1}=\|\beta_{\Sp}\|_{1}+\|\beta_{\Sp^{\perp}}\|_{1}$
for any $\Sp\subset\{1,\ldots,p\}$ and $\Sp^{\perp}=\{1,\ldots,p\}\backslash\Sp$.
If we assume the true $\beta$ is $\mathcal{S}$-sparse, i.e. $\beta_{\mathcal{S}^{\perp}}=0$,
then applying Holder's inequality to the right of Eq.~\ref{eq: rev3}
gives us the upper bound
\begin{equation}
\frac{1}{n}\lVert X\hatDelta\rVert_{2}^{2}\le\frac{2}{n}\lVert X^{\top}Lu\rVert_{\infty}\lVert\hatDelta\rVert_{1}+2\lambda_{n}\{\lVert\beta_{\Sp}\rVert_{1}-\lVert\beta_{\Sp}+\hat{\Delta}_{\Sp}\rVert_{1}-\lVert\hat{\Delta}_{\Sp^{\perp}}\rVert_{1}\}\;.\label{eq:lasso_general_upper_bound}
\end{equation}
To further control the right-hand-size of this bound, we now make
a convenient choice for $\lambda_{n}$ that upper bounds the projection
of the errors onto the columns of our design matrix. We will define
this choice as the event
\begin{equation}
\mathcal{E}_{\lambda}:=\big\{\frac{2}{n}\|X^{\top}Lu\|_{\infty}<\lambda_{n}\big\}
\end{equation}
conditional on which, we obtain the bound 
\begin{equation}
n^{-1}\lVert X\hatDelta\rVert_{2}^{2}\le\lambda_{n}(3\|\hat{\Delta}_{\mathcal{S}}\|_{1}-\|\hat{\Delta}_{\Sp^{\perp}}\|_{1})\le3\lambda_{n}\sqrt{|\mathcal{S}|}\|\hat{\Delta}\|_{2}\;.\label{eq:lasso_lambda_ub}
\end{equation}
To decouple $\|\hat{\Delta}\|_{2}$ from the design matrix on the
left-hand-side of Eq.~\ref{eq:lasso_lambda_ub} we may consider the
minimum eigenvalue $\min_{z\in\mathbb{R}^{p}}n^{-1/2}\|Xz\|_{2}/\|z\|_{2}$.
However, in the case of the LASSO, we do not need to consider all
directions $z\in\mathbb{R}^{p}$, but rather only within a cone
\begin{equation}
\mathbb{C}_{\alpha}(\Sp):=\left\{ z\in\mathbb{R}^{p}\;|\;\|z_{\Sp^{\perp}}\|_{1}\le\alpha\|z_{\Sp}\|_{1}\right\} \;.\label{eq:calpha}
\end{equation}
Intuitively, the definition of this cone allows for us to bound the
size of the vector on the out-of-subspace ($\Sp^{\perp}$) components
based on those within the model sub-space. Considering the lower-bound
in (\ref{eq:lasso_lambda_ub}) we find $3\|\hat{\Delta}_{\mathcal{S}}\|_{1}-\|\hat{\Delta}_{\Sp^{\perp}}\|_{1}\ge0$,
leading to interest vectors in the cone $\mathbb{C}_{\alpha}(\mathcal{S})$,
with $\alpha=3$ . We say that $X$ satisfies the \emph{restricted
eigenvalue (RE)} condition over $\Sp$, with $\kappa,\alpha$ if the
event:
\begin{equation}
\mathcal{E}_{\mathrm{RE}}(n^{-1/2}X;\kappa,\alpha):=\left\{ n^{-1/2}\|Xz\|_{2}\ge\kappa\|z\|_{2}\quad\forall z\in\mathbb{C}_{\alpha}(\Sp)\;\right\} \;\label{eq:RE}
\end{equation}
holds. Conditional on $\mathcal{E}_{\mathrm{RE}}(n^{-1/2}X;\kappa,\alpha)$
and $\mathcal{E}_{\lambda}$, then re-arranging Eq.~\ref{eq:lasso_lambda_ub}
gives the bound
\[
\|\hat{\Delta}\|_{2}\le3\frac{\lambda_{n}}{\kappa^{2}}\sqrt{|\mathcal{S}|}\;,
\]
where we note the crucial importance of the term $\kappa$ and the
scaling of the regularization parameter $\lambda_{n}$. 

\subsection{The AR($q$) Example}

Dependence in the errors $e$ will not impact the RE condition, which
only depends on the design. As such, we will first examine how serial
dependence in $e$ impacts the choice of $\lambda_{n}$. In particular,
we will look to specify a $\lambda_{n}$ that enables $\mathcal{E}_{\lambda}$
to hold in high probability. We will work in a setting which allows
$p\gg n$, and (typically) develop our bounds to hold in probability
related to $p$, such that as $p\rightarrow\infty$ our results will
hold with probability one. We characterize convergence in terms of
the decay function
\[
\delta(n,r):=\sqrt{\frac{\log r}{n}}\;,
\]
which for consistency we require $\delta(n,p^{\tau})\rightarrow0$.
Our analysis will take place in a sub-Gaussian setting, that is, we
assume both the design matrix and the error process are sub-Gaussian.
We formalize this assumption below:

\begin{assumption}{Sub-Gaussian Analysis}\label{ass:1}

Let $X=W\Sigma^{1/2}$, where $W$ is an isotropic $n\times p$ dimensional
sub-Gaussian random matrix, and thus $\Sigma$ represents the covariance
across columns of $X$. Specifically, we assume that the projection
of these variables over rows/columns is also a sub-Gaussian variable,
with norm
\begin{align*}
\sup_{\|v\|_{2}=1}\normiii{u^{\top}v}_{2} & \le K_{u}\\
\sup_{\|v\|_{2}=\|z\|_{2}=1}\normiii{z^{\top}Wv}_{2} & \le K_{W}
\end{align*}
for some absolute constants $K_{W},K_{u}$, not dependent on $n,p$.

\end{assumption}

We note that as a consequence of the above assumption, $X$ will also
be sub-Gaussian such that we can find a constant $K_{X}$ whereby
$\sup_{\|v\|_{2=1}}\normiii{W\Sigma^{1/2}v}_{2}\le\max_{i}\|\Sigma_{\cdot,i}^{1/2}\|_{2}K_{W}=K_{X}$.
Thus, whilst $K_{W}$ is not a function of $n,p$, we see that $K_{X}$
may be in the case where the covariates exhibit persistent cross-correlation.
For the purposes of this work, as our interest is in primarily correcting
for autocorrelation, we assume that a finite $K_{X}$ can be found
that holds for all $n,p$.

\begin{figure}
\begin{centering}
\includegraphics[width=0.7\columnwidth]{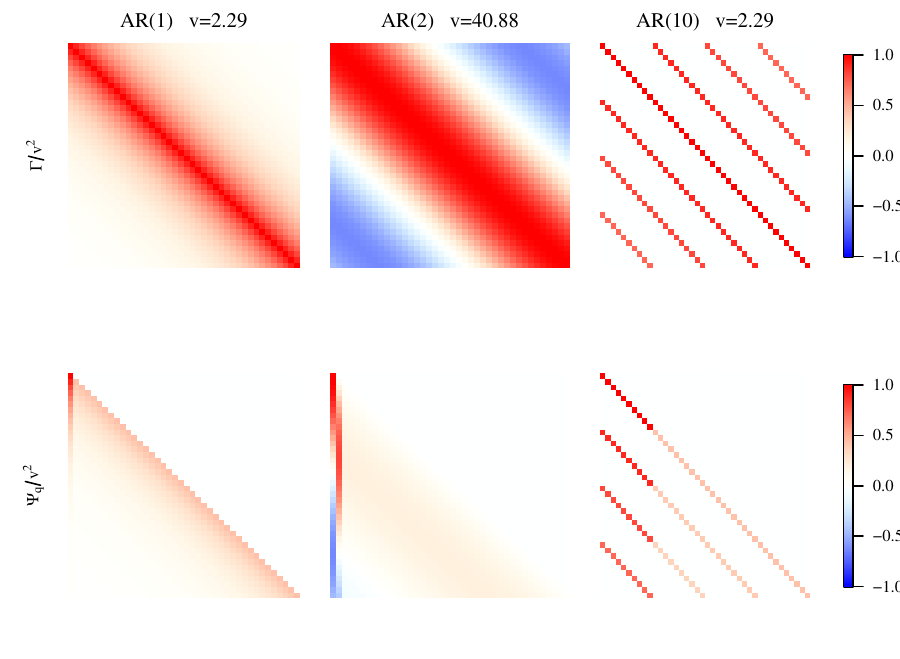}
\par\end{centering}
\caption{Heatmaps of $\Gamma$ (top) and $\Psi$ (bottom) corresponding to
various stationary AR processes, with $\sigma_{u}^{2}=1$. Left: AR(1)
$\phi_{1}=0.9$; Middle: AR(2) $\phi=(1.96,-0.97)$; Right: AR(10)
$\phi_{j}=0$ for $j=1,\ldots,9$ and $\phi_{10}=0.9$. Note: whilst
$\Gamma$ is always Toeplitz, the corresponding $\Psi$ is Toeplitz
only for columns $j>q$.\label{fig:arp_L}}

\end{figure}

To study in detail how dependence in the errors impacts the quantity
$X^{\top}Lu$ we study the case where $L$ is associated with an AR($q$)
model. That is, we assume the errors follow the model
\begin{equation}
e_{t}=\phi_{1}e_{t-1}+\ldots+\phi_{q}e_{t-q}+u_{t}\;,\label{eq:arp}
\end{equation}
for $t\in\mathbb{Z}$, and given parameters $\{\phi_{i}\}_{i=1}^{q}$.
For simplicity and analytical tractability, we assume here that $q$
is fixed as $n,p$ grow, however, the arguments we put forward can
feasibly be expanded to allow $q$ (or indeed the specific parameters
$\phi_{i}$) to be a sequence in $n$, and thus use the AR($q$) to
approximate any covariance stationary error process, c.f. \cite{Goldenshluger2001}.
We recall that for a stationary (causal) AR($q$) parameterization,
the parameters $\phi=(\phi_{1},\ldots,\phi_{q})^{\top}$ induce a
Toeplitz autocovariance matrix for the sequence of points indexed
$t=1,\ldots,n$. In particular, we are interested in the Cholesky
decomposition of this matrix, that is $L=\sigma_{u}^{-1}\Gamma^{1/2}$
where $\Gamma=\sigma_{u}^{2}LL^{\top}$. For the AR($q$) case, we
denote these $n\times n$ lower-triangular Cholesky matrices as $\Psi_{q}$.
If $q=0$ then we have a white-noise error process, and $\Psi_{0}=I_{n}$,
if $q=1$, then we have dependent errors with the level of persistence
dictated by $\phi\in(-1,1)$, the matrix $L=\Psi_{1}$ takes the form

\[
\Psi_{1}=\begin{pmatrix}v & 0 & \cdots & 0 & 0\\
v\phi & 1 & \cdots & 0 & 0\\
\vdots & \vdots & \ddots & \vdots & \vdots\\
v\phi^{n-1} & \phi^{n-2} & \cdots & \phi & 1
\end{pmatrix}\;.
\]
where $v^{2}=\mathrm{Var}[e_{t}]/\sigma_{u}^{2}$ is the variance
of the process after adjusting for the scale of the driving white
noise process. In the AR(1) case we have $v=(1-\phi^{2})^{-1/2}$.
For $q\ge2$ the form of this matrix becomes more complex, however,
can be determined via the Yule-Walker equations \cite{Brockwell2009}.
In general, we find that the first column is the scaled autocorrelation
function, however, the structure in other columns is modified, until
for columns $j>q$, where we obtain a regular Toeplitz form. Some
examples of the autocovariance $\Gamma$ and corresponding forms for
$\Psi_{q}$ can be seen in Figure~ \ref{fig:arp_L}.

The added flexibility of the AR($q$) model beyond the $q=1$ case
is useful, in that these matrices can represent a large class of covariance
stationary error processes, however, makes the interpretation of theoretical/empirical
results more complicated. As such, we will focus on deriving results
that hold for the AR($q$) error process, however, will provide discussion
in relation to the AR(1) setting where we have a known (simple) form
for $L=\Psi_{1}$, and variance $v^{2}$. We start our analysis with
some foundational deviation bounds that utilize our sub-Gaussian assumptions.

\begin{lemma}\label{lemma:tail}Under Assumption \ref{ass:1} we
have
\begin{equation}
P\left[\|X^{\top}Lu\|_{\infty}\ge\epsilon\right]\le2p\exp\left[-c\min\left(\frac{\epsilon^{2}}{K_{X}^{2}K_{u}^{2}\lVert L\rVert_{F}^{2}},\frac{\epsilon}{K_{X}K_{u}\lVert L\rVert_{2}}\right)\right]\;,\label{eq:bi-linear}
\end{equation}
where $c>0$ is an absolute constant.

\end{lemma}

This result is similar to that of the Hanson-Wright bound (c.f. \cite{vershynin2018high})
for the sub-Gaussian chaos $u^{\top}Au$. However, in our case, we
have a bi-linear form $u^{\top}Au'$ where the variables either side
of the matrix are both sub-Gaussian, but not the same variable. Similar
bounds can also be found in the work of Park et al. \cite{Park2021,Park2023}.
Let us now recall the event $\mathcal{E}_{\lambda}:=\{\frac{2}{n}\|X^{\top}Lu\|_{\infty}<\lambda_{n}\}$,
associating $\lambda_{n}$ with $\epsilon$ in (\ref{eq:bi-linear})
we see that for $\mathcal{E}_{\lambda}$ to hold in high-probability
we should account for the magnitude of the matrix $L$, via $\|L\|_{F}^{2}$
or $\|L\|_{2}$. If the matrix $L$ is large, then we will need to
have a correspondingly larger $\lambda_{n}$ to ensure $\mathcal{E}_{\lambda}$
holds for some fixed probability. For a covariance stationary error
process, this scale is given in relation to the variance and we find
$\|L\|_{F}^{2}=nv^{2}$ and $\|L\|_{2}=v$. 

\begin{corollary}{Thresholding with AR($q$) errors}\label{cor:ar_deviation_bound}

Under Assumptions \ref{ass:1}, then for all\footnote{Our choice of $\delta(n,p^{\tau})$ allows us to use the Gaussian
tail decay for a sufficient sample size. If this sample size isn't
met in practice, then one could consider the sub-Exponential tail
instead (second term in the minima for the bound in Lemma \ref{lemma:tail}). } $n\ge2^{-2}c\tau\log p$ and $\tau>1$, if we set 
\begin{equation}
\lambda_{n}=\frac{K_{X}K_{u}}{c^{1/2}}v\delta(n,p^{\tau})\;,\label{eq:lambda_choice}
\end{equation}
then we have $P[\mathcal{E}_{\lambda}]\ge1-2p^{1-\tau}$ for some
absolute constant $c>0$.

\end{corollary}

From this choice of $\lambda_{n}$, we see that if $\mathcal{E}_{\mathrm{RE}}(n^{-1/2}X;\kappa,3)$
held over $\mathcal{S}_{0}=\mathrm{supp}(\beta)$, then we would obtain
the bound
\begin{equation}
\|\hat{\Delta}\|_{2}\le3\frac{K_{X}K_{u}}{c^{1/2}\kappa}v\sqrt{\frac{s\log p^{\tau}}{n}}\label{eq:lasso_l2_err}
\end{equation}
where $s=|\mathcal{S}_{0}|$. The rate above is the same as one would
expect if the errors were independent (e.g. see \cite{Negahban2012,Wainwright2009,wainwright2019high}),
with the only difference being the variance of the error process,
i.e. the bound is proportional to $v$. Correspondingly, if $v$ is
very large, which can happen in the highly-persistent setting, then
we would expect the errors incurred by the LASSO to be considerable.

\begin{figure}
\centering
\includegraphics[width=0.8\columnwidth]{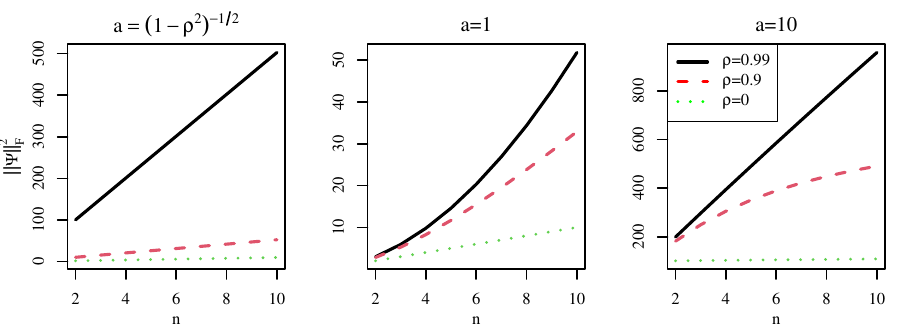}\caption{Comparison of the growth in $\|L\|_{F}$ for an AR(1) process as a
function of $n$, for different $\phi_{1}$ and initial variances $v_{0}$.
\label{fig:frobenius_terms}}
\end{figure}

We note that the bound in Lemma \ref{lemma:tail} can still be applied
even when the errors are not stationary. For example, in the autoregressive
setting one may consider the situation where the initial variance
for the error was not that of the stationary variance, i.e., we may
have $\mathrm{Var}[e_{1}]\ne v^{2}\sigma_{u}^{2}$. In this case,
there will be a period of adaption in the variance of the errors as
$n$ increases. For instance, if we set $\mathrm{Var}[e_{1}]=\sigma_{u}^{2}$,
then the term $\|L\|_{F}^{2}$ would initially grow as a quadratic
in $t$ until $t>\min_{t}(\mathrm{Var}[e_{t}]=v^{2}\sigma_{u}^{2})$
after which it becomes linear, some examples of this behavior can
be seen in Fig. \ref{fig:frobenius_terms}. Although not typical in
time-series applications, this setting may prove relevant if one assumes
the regression errors compound from a known starting distribution.
This case also relates to the so-called local-to-unity framework \cite{phillips1987towards}
where one lets $\rho_{n}=\exp(\epsilon/n)$ and $\rho_{n}\rightarrow1$
when the non-centrality parameter $\epsilon$ is kept fixed at $\epsilon\approx0$,
as $n\to\infty$ . In such a setting, it would be important to track
not just the scale, but also the rate (in $n$) of the $\|L\|_{F}^{2}$
term. A further setting one can consider, is that where there is some
prior knowledge of how the variance of the errors fluctuates, this
can then be adjusted for by considering the scaling of $\|L\|_{F}$
and $\|L\|_{2}$. Although all these directions can be of interest,
we focus in this paper on the stationary AR($q$) error setting.

\subsection{Empirical Behavior}

\begin{figure}
\centering
\includegraphics[width=0.8\columnwidth]{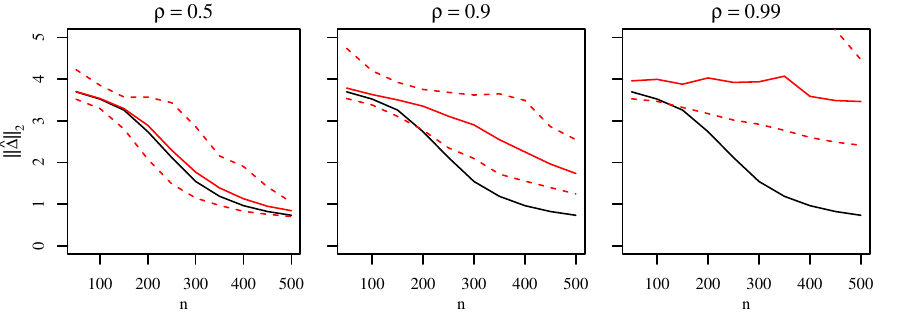}\caption{Estimation error $p^{-1/2}\|\hat{\Delta}\|_{2}$ achieved by the LASSO
with varying AR($1$) error parameter. Black line indicates performance
in independent error setting $\sigma_{u}^{2}=1$. Solid and dashed
lines represent the mean and 95\% confidence intervals respectively,
based on 1000 simulations.\label{fig:lasso_exp}}
\end{figure}

To illustrate the behavior of the LASSO with autocorrelated errors,
here we provide a simple synthetic study in the AR(1) setting. The
coefficient $\phi_{1}$ is varied from $0$ to very near $1$. In
this case, both the design matrix $X$ and the errors $e$ are simulated
as Gaussian random-variables, with the former being isotropic in nature.
To investigate the specific impact of the autocorrelation, we contrast
the performance of the LASSO with AR errors to that of \emph{i.i.d.}
errors, and let $\sigma_{u}^{2}=1$. The results of this experiment
are given in Figure \ref{fig:lasso_exp}, where we see that the performance
of the LASSO, degrades both as a function of the overall error variance,
and the persistence in the autocorrelation function. Perhaps the most
surprising observation here is that for moderate levels of persistence
$\phi=0.5$, the performance of the LASSO is not too degraded relative
to the independent errors (i.e., given by the black lines in the figure).
However, overall, the theoretical and empirical results combine to
illustrate the negative impact of autoregressive errors in finite
samples. 

\section{Regularized Generalized Least Squares}

\label{sec:rgls}

\subsection{GLS-LASSO}

Let us suppose (for now) that the matrix $L$ is known a-priori. We
can immediately obtain a whitening matrix $R=L^{-1}$ and thus construct
the regression $\ty=\tX\beta+u$, with $\ty=Ry$, $\tilde{X}=RX$
and $\mathrm{Cov}(u)=I_{n}$. We consider the GLS-LASSO estimator
of the form
\begin{equation}
\tbeta=\arg\min_{\beta\in\mathbb{R}^{p}}\bigg(\frac{1}{2n}\lVert\ty-\tX\beta\rVert_{2}^{2}+\tilde{\lambda}_{n}\lVert\beta\rVert_{1}\bigg)\;,\label{eq: lasso2}
\end{equation}
where $\tilde{\lambda}_{n}\ge0$ is distinct from $\lambda_{n}$.
Whilst we will show that $\tilde{\lambda}_{n}$ can be chosen to be
smaller than $\lambda_{n}$, resulting in a lower error bound, we
still need to verify the RE conditions hold on the rotated design.
Previously, in settings not relative to time-series \cite{Jia2015}
demonstrated that such rotation was undesirable as it destabilised
the design and made the RE condition unlikely to occur, or indeed
impossible. Our setting is significantly different, as we have a very
particular form of rotation matrix that we are applying. 

\begin{figure}
\begin{centering}
\includegraphics[width=0.7\columnwidth]{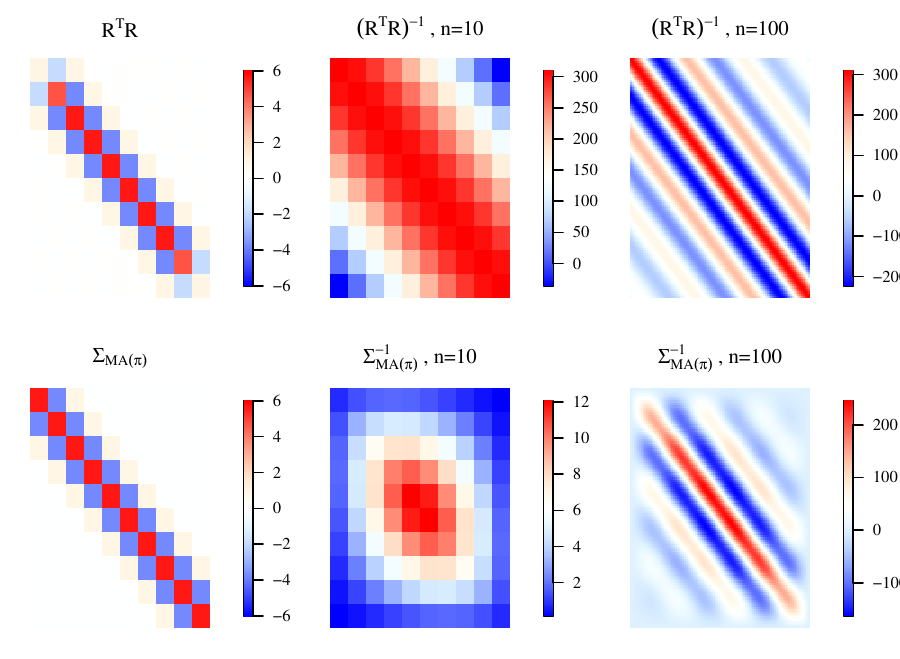}
\par\end{centering}
\caption{Comparison of $R^{\top}R$ and $\Sigma_{\mathrm{MA}(\pi)}$ (and their
inverses) obtained when $e_{t}$ is generated by an AR(2) model with
$\phi_{1}\approx2$, $\phi_{2}\approx-1$, i.e. on the edge of the
stationary regime. Left and middle plots are generated for $n=10$
samples to highlight the discrepancy at the boundary, whilst the figures
on the right represent the same plots for $n=100$ showing good approximation
to the stationary autocovariance associated with this AR(2) model.\label{fig:RTR approx}}
\end{figure}

To this end, this section specializes on the setting where $L=\Psi_{q}$
corresponding to the AR($q$) error model. This choice allows us to
obtain specific forms of whitening matrix, for example, if one inverts
the AR($q)$ matrix $\Psi_{q}$ we obtain\textcolor{black}{
\begin{equation}
R=\Psi_{q}^{-1}=\left(\begin{array}{ccccc}
{\color{black}{\color{blue}}} & {\color{black}{\color{blue}R_{I}}} & {\color{black}{\color{blue}\vdots}} & 0\\
{\color{black}{\color{blue}\cdots}} & {\color{black}{\color{blue}\cdots}} & {\color{black}\mathclose{\color{blue}\lrcorner}} & 0 & 0\\
-\phi_{q} & \cdots & -\phi_{1} & 1 & 0\\
0 & -\phi_{q} & \cdots & -\phi_{1} & 1
\end{array}\right)\;\label{eq:Rform}
\end{equation}
}where the first block $R_{I}\in\mathbb{R}^{q\times q}$ (in blue
above) is distinct from the regular Toeplitz pattern (for rows $q+1,\ldots,n$)
which take the well known form based on the AR($q$) coefficients
\cite{Brockwell2009}. Similar to the importance of $\|L\|_{F}$ and~$\|L\|_{2}$
when determining the inflation of the $\lambda_{n}$ due to dependence
in the errors, in this case we will be interested in the scale of
$\|R\|_{F}$. In the case where $R=\Psi_{q}^{-1}$ we see that $R$
will induce a moving-average like autocovariance structure in $\tilde{X}$,
a link which we formalize below.

\begin{lemma}\label{lemma:ma_comparison}

Let us consider the MA($q$) model induced by $\pi=(1,-\phi_{1},\ldots,-\phi_{q})$
with autocovariance given as $\gamma_{|i-j|}=\sum_{k=0}^{q-|i-j|}\pi_{k}\pi_{k+|i-j|}$.
Let $[\Gamma_{\mathrm{MA}(\pi)}]_{ij}:=\gamma_{|i-j|}$ for $i,j\in1,\ldots,n$.
For any stationary AR($q$) model with $q\le2$, we have
\begin{equation}
\|R^{\top}R\|_{2}\le\|\Sigma_{\mathrm{MA}(\pi)}\|_{2}\le(2q+1)\|\pi\|_{2}^{2}\;,\label{eq:RtRapprox}
\end{equation}
furthermore, we have $\|R\|_{2}\le\|\pi\|_{1}\le(q+1)\|\pi\|_{2}$.\end{lemma}

We conjecture this result holds for $q$ of higher order, however,
the proof requires verification over the stationary regimes for higher
order models, which is not immediately obvious. In the following,
we will proceed assuming the condition (\ref{eq:RtRapprox}) holds.
The similarity between $\Sigma_{\mathrm{MA}(\pi)}$ and $R^{\top}R$
is illustrated in Figure \ref{fig:RTR approx}.

We are now in a position to assess the impact of the rotation matrix
on the design. Our first result in this direction extends previous
work \cite{Rudelson2012} by applying a reduction principle that ensures
the RE condition will hold on the rotated design (in high probability)
if an RE condition (with tighter conditions) holds on the population
covariance. 

\begin{proposition}{RE Condition for Rotated Design (Sub-Gaussian)}

\label{prop:re_subgaussian}

Assume the RE condition $\mathcal{E}_{\mathrm{RE}}(\Sigma^{1/2};\kappa,3\alpha)$
holds for all $\mathcal{S\subset}[p]$ such that $|\mathcal{S}|=s<p$.
Let $\zeta$ and $c$ be positive constants\footnote{Note this $c$ is a different constant to those given earlier, e.g.
in Lemma \ref{lemma:tail}. Given the complexity of the bounds, in
this section, these generic constants should not be compared between
results. If there are multiple distinct constants in a result, these
will be labeled $c_{1},c_{2},\ldots$, however, again these should
not be considered the same across results. We make an effort to track
key quantities, e.g. $K_{W},\pi,q$ etc contributing to these constants,
but other terms are encapsulated by generic constants to simplify
the presentation.} such that 
\[
q/n<\zeta<\min[1,10K_{W}^{2}\|\pi\|_{2}(q+1)]\;.
\]
Then given sufficient samples
\begin{equation}
n\ge\frac{K_{W}^{4}\|\pi\|_{2}^{2}(q+1)^{2}\alpha^{3}}{c}\max_{j\in\mathcal{S}}\|\Sigma_{\cdot,j}^{1/2}\|_{2}^{2}\frac{s}{\kappa^{2}\zeta^{4}}\log(p^{\tau})\;,\label{eq:sample_size_subg}
\end{equation}
we have the RE condition $\mathcal{E}_{\mathrm{RE}}(n^{-1/2}\tilde{X};\tilde{\kappa},\alpha)$
holds for some $\tilde{\kappa}\ge(1-\zeta)\|\pi\|_{2}\kappa$ with
probability greater than $1-2p^{-\tau}$.

\end{proposition}

Our proof extends the previous arguments of \cite{Rudelson2012} by
using the Hanson-Wright bound (Lemma \ref{lemma:hanson_wright}, c.f.
\cite{vershynin2018high}) to account for the impact of $R$ which
induces dependence across the rows of $\tilde{X}$. We see that the
impact of $R$ is to scale the eigenvalues by $\|\pi\|_{2}(1-\zeta)$
compared to the population values of $\kappa$. It should be noted,
that the sample size requirement (\ref{eq:sample_size_subg}) presented
above is a simplification (that can be seen as been sufficient, but
not necessary) and we see this is inflated when one chooses a small
$\zeta$, therefore taking $\zeta\rightarrow0$ requires an increasing
number of samples; this minimum sample size also scales linearly with
the size of the support $s$ as one may expect. 

Whilst the arguments above for the sub-Gaussian design are useful
in that they provide statistical guarantees under a wide range of
design choices, the bounds and constants involved above are not explicit.
We now look at a more specific case where we assume the design matrix
is Gaussian, which allows for arguably finer control of the lower
bound on the eigenvalues. The argument is based on the paper by Raskutti
et al. \cite{Yu2010}, which we generalize to the setting where rows
of the design matrix can be correlated, in our case due to the rotation
matrix $R$. We also provide explicit constants in the probability
statements afforded by the Gaussian assumption.

\begin{lemma}{RE Condition for Rotated Design (Gaussian)}\label{lemma:raskutti}

For $\tilde{X}=RX$ with $X\sim\mathcal{N}_{n\times p}(0;I_{n},\Sigma)$,
let $\omega$ be a positive constant such that $q^{3}/n\le\omega<1$,
then we have
\[
\frac{1}{\sqrt{n}}\frac{\|\tilde{X}v\|_{2}}{\|\pi\|_{2}}\ge(1-\omega)\|\Sigma^{1/2}v\|_{2}-9\|\Sigma\|_{\infty}\delta(n,p)\|v\|_{1}\quad\forall v\in\mathbb{R}^{p}
\]
in probability at least $1-3\exp(-2^{-1}\omega n)$. 

\end{lemma}

As a corollary of the above, we have that for $\|\Sigma^{1/2}v\|_{2}\ge\kappa\|v\|_{2}$,
i.e. if an RE condition $\mathcal{E}_{\mathrm{RE}}(\Sigma^{1/2};\kappa,\infty)$
occurs over $\mathbb{R}^{p}$, then we have high confidence that the
rotated design will satisfy an RE condition $\mathcal{E}_{\mathrm{RE}}(n^{-1/2}\tilde{X};(1-\omega)\kappa\|\pi\|_{2},\alpha)$
as long $9\|\Sigma\|_{\infty}\delta(n,p)\|v\|_{1}\le2^{-1}(1-\omega)\kappa\|v\|_{2}$.
This latter condition puts a constraint on the size of $\mathcal{S}$
which the rotated design will satisfy the RE condition over, specifically
\begin{equation}
|\mathcal{S}|\le\frac{(1-\omega)^{2}\kappa^{2}}{3^{4}2^{2}\|\Sigma\|_{\infty}^{2}}\frac{n}{\log p}\;.\label{eq:sparsity_gaussian}
\end{equation}

One can equally view this bound on the support set, as a minimum sample
size requirement, given a fixed $s=|\mathcal{S}$|. Comparing with
the sample size requirement in the sub-Gaussian case (\ref{eq:sample_size_subg}),
we note there is no impact of the rotation, e.g. no $\|\pi\|_{2}$
terms appear in Eq. \ref{eq:sparsity_gaussian}. To this end, we observe
that the rotation matrix $R$ does have an effect compared to the
independent case; however, this impact is in the magnification of
eigenvalues and is minimal, even for highly persistent AR($q$) processes,
due to the fact $\|\pi\|_{2}$ is typically bounded by a small constant.
A further point of comparison surrounds $\omega$, which plays a similar
role to $\zeta$ in measuring the reduction of the eigenvalues, however,
in this bound we find the minimum sample size scales as $1/\omega^{2}$
as opposed to $1/\zeta^{4}$ in the sub-Gaussian case. Given the more
precise characterization of the eigenvalue reduction by explicitly
accounting for $\|v\|_{1}$ in the Gaussian case we now construct
an oracle inequality that allows one to assess contribution to the
GLS estimation error from both on-support and off-support terms, i.e.
a bound which covers the situation where $\mathcal{S}$ may include
elements not in the true support $\mathcal{S}_{0}$.

\begin{proposition}{GLS-LASSO Oracle Inequality}

\label{prop:oracle}

Let $\tilde{\Delta}:=\tilde{\beta}-\beta_{0}$, and set $\tilde{\lambda}_{n}\ge2c^{-1/2}K_{u}\|\Sigma^{1/2}\|_{2}\|\pi\|_{2}\delta(n,p^{\tau})$
for $\tau>1$ where $c>0$ is an absolute constant. Assume that $X\sim\mathcal{N}_{n\times p}(0,I_{n};\Sigma)$
with $\|\Sigma^{1/2}v\|_{2}\ge\kappa\|v\|_{2}/\|\pi\|_{2}$ and errors
$\{e_{t}\}$ are generated by a sub-Gaussian AR($q$) process. Let
$\tilde{\kappa}=(1-\omega)\kappa$ with $q^{2}/n<\omega<1$, then
for
\[
n\ge\max\left[\frac{3^{4}2^{6}\|\Sigma\|_{\infty}|\mathcal{S}|}{\tilde{\kappa}^{2}},\frac{(q+1)^{2}}{c}\right]\log(p^{\tau})\;.
\]
we have
\[
\|\hat{\Delta}\|_{2}^{2}\le\underset{\mathrm{estimation\:error}}{\underbrace{\frac{2^{3}.3^{2}}{\tilde{\kappa}^{4}\lVert\pi\rVert_{2}^{4}}\tilde{\lambda}_{n}^{2}|\mathcal{S}|}}+\underset{\mathrm{approximation\:error}}{\underbrace{\frac{2^{2}}{\tilde{\kappa}^{2}\lVert\pi\rVert_{2}^{2}}\tilde{\lambda}_{n}\lVert\beta_{\mathcal{S}^{\perp}}\rVert_{1}+\frac{2^{6}3^{4}\|\Sigma\|_{\infty}^{2}}{\tilde{\kappa}^{2}}\delta^{2}(n,p)\lVert\beta_{\mathcal{S}^{\perp}}\rVert_{1}^{2}}}\;,
\]
with probability at least $1-2p^{1-\tau}-3\exp(-2^{-1}\omega n)$.

\end{proposition}

Overall, with the minimal choice of $\tilde{\lambda}_{n}$, we have
$\|\Delta\|_{2}^{2}=\mathcal{O}_{P}(|S|\log p/n)$ where $\mathcal{S}_{0}\subseteq\mathcal{S}$.
This mimics the earlier bound of Eq. \ref{eq:lasso_l2_err}, however,
the bound allows the choice $\tilde{\lambda}_{n}=\lambda_{n}\|\pi\|_{2}/v$,
so the factor of $v$ due to temporal dependence is removed in the
GLS case. The oracle inequality also allows us to assess the impact
of components in $\mathcal{S}\setminus\mathcal{S}_{0}$ where this
impact is measured by terms involving $\lVert\beta_{\mathcal{S}^{\perp}}\rVert_{1}$
contributing to the \emph{approximation error} component highlighted
above. We see the terms involved here are bounded to zero asymptotically
since both $\tilde{\lambda}_{n}$ and $\delta^{2}(n,p)\rightarrow0$.
The estimation error is controlled as $\tilde{\lambda}_{n}\rightarrow0$,
where the inflationary factor of $\|\pi\|_{2}$ in the choice of $\tilde{\lambda}$
cancels to some extent the scaling of the smallest eigenvalues found
in Lemma \ref{lemma:raskutti}. Finally, if one desired, we could
choose $\omega$ as a sequence in $n$ as long as it satisfies the
condition $2q^{3}/n\le\omega$, this would permit a choice $\omega\asymp\delta^{2}(n,p)$
allowing $(1-\omega)\kappa\rightarrow\kappa$ whilst still maintaining
the bound in probability. In essence, if the rotation matrix is correctly
specified, the GLS-LASSO achieves the same performance as would a
regular LASSO faced with uncorrelated errors, even in finite samples.

Of course, this version of the GLS has assumed exact knowledge of
the whitening matrix, which in practice is not available to us, we
consider a feasible version of the estimator in the following section.

\subsection{Feasible GLS-LASSO}

To enable a feasible GLS procedure, we first require an estimate of
the AR parameter based on an estimated residual series. Let us consider
the estimated residuals, based on an estimator of $\hat{\beta}$,
to be defined as $\hat{e}_{t}=y_{t}-\hat{\beta}^{\top}x_{t}=e_{t}+\hat{\Delta}^{\top}x_{t}$.
We propose to use the OLS estimator for the AR coefficients based
on the VAR(1) representation of AR($q$), i.e.
\begin{equation}
\hat{\phi}=(\hat{E}^{\top}\hat{E})^{-1}\hat{E}^{\top}\hat{e}'\label{eq:OLS}
\end{equation}
where the ($n-q)\times q$ dimensional matrix 
\[
\hat{E}=\left(\begin{array}{ccc}
\hat{e}_{q} & \ldots & \hat{e}_{1}\\
\vdots &  & \vdots\\
\hat{e}_{n-1} & \ldots & \hat{e}_{n-q}
\end{array}\right)\;,
\]
is constructed from stacking lagged values of $\{\hat{e}_{t}\}_{t=1}^{n}$,
and $\hat{e}'=(\hat{e}_{q+1},\ldots,\hat{e}_{n})^{\top}$. There have
been a couple of recent works looking at finite sample properties
of OLS estimation for autoregressive processes, e.g. \cite{bercu2008exponential}
in the AR(1) setting, and \cite{Gonzlez2020} in the AR($q$) setting,
however, both of these works are based on observing the true realizations
of the process, not some noisy version potentially contaminated via
the impact of covariates. Our result below can thus be of independent
interest, where our approach utilizes the same deviation bounds for
quadratic forms (e.g. Lemmas \ref{lemma:tail} and \ref{lemma:hanson_wright})
used in the previous arguments to control the random matrices involved
in Eq. \ref{eq:OLS}, whilst conditioning on $\|\hat{\Delta}\|_{2}$
of a limited size.

\begin{proposition}{Feasible AR(q) Errror}\label{prop:phi_err}

Under Assumption \ref{ass:1} and that we have auto-regressive sub-Gaussian
errors of order $q$ and the errors incurred by the first stage LASSO
$\hat{\beta}$ are bounded according to $\|\hat{\Delta}\|_{2}\le K_{u}/K_{x}$,
then with sufficient samples
\[
n\ge\max\left[1+q,\frac{K_{u}^{4}}{\sigma_{u}^{4}}q^{4}\frac{\gamma_{\max}^{2}(\Gamma)}{\gamma_{\min}^{2}(\Gamma)}\right]\frac{\log(p^{\tau})}{c_{1}}
\]
we have the error bound

\begin{equation}
\|\hat{\phi}-\phi\|_{\infty}\le\frac{2q^{2}}{\gamma_{\min}(\Gamma_{q})}\left[\|\Sigma^{1/2}\Delta\|_{2}^{2}\|\phi\|_{\infty}+2c_{2}^{-1/2}K_{u}^{2}v\delta(m,p^{\tau})\right]\label{eq:phi_bound}
\end{equation}
in probability greater than $1-14p^{1-\tau}$ , where $c_{1},c_{2}$
are distinct positive constants.

\end{proposition}

This bound is developed conditional on the performance of the first-stage
LASSO meeting a set criteria, that is, if the initial error incurred
by the LASSO is not too great, then the estimation error for the AR
coefficients is controlled as one may expect, i.e. proportional to
$v/n^{1/2}$. As a corollary of Prop. \ref{prop:phi_err} and using
$\|\Delta\|_{2}^{2}=\mathcal{O}_{p}(v^{2}s\delta^{2}(n,p^{\tau}))$
from (\ref{eq:lasso_l2_err}) we obtain the following result on the
feasible errors.

\begin{corollary}\label{cor:phi_bound}

Under Assumption \ref{ass:1} and that $\mathcal{E}_{RE}(\Sigma^{1/2},\kappa,3\alpha)$
occurs for $\mathcal{S}=\mathcal{S}_{0}$, $|\mathcal{S}_{0}|=s$
and $\alpha=3$; let $\lambda_{n}=c_{1}^{-1/2}K_{X}K_{u}v\delta(n,p^{\tau})$
be chosen as per Corollary \ref{cor:ar_deviation_bound}. Then given
sufficient samples
\[
n\ge c_{2}\kappa^{-2}q^{4}v^{2}s\max[1,\kappa^{-2}\|\Sigma^{1/2}\|_{2}^{4}\|\phi\|_{\infty}^{2}s]\log(p^{\tau})
\]
we have
\begin{equation}
\|\hat{\phi}-\phi\|_{\infty}\le c_{3}\frac{q^{2}K_{u}^{2}v}{\gamma_{\min}(\Gamma_{q})}\delta(n,p^{\tau})\;,
\end{equation}
in probability at least $1-18p^{1-\tau}$, where $c_{1},c_{2},c_{3}$
are distinct positive constants.

\end{corollary}

The sample size requirement here is a combination of those required
for the RE condition, and for the convergence of the AR error bound.
Specifically, we require a sample size that ensures $\|\hat{\phi}-\phi\|_{\infty}=\mathcal{O}(\delta(n,p^{\tau}))$,
a rate which is limited by the second term in Eq. \ref{eq:phi_bound}
and requires the minimum sample grows as a quadratic in sparsity $s$.

We now consider the FGLS estimator defined as
\begin{equation}
\bar{\beta}=\arg\min_{\beta\in\mathbb{R}^{p}}\left[\frac{1}{2n}\|\hat{R}(y-X\beta)\|_{2}^{2}+\bar{\lambda}_{n}\|\beta\|_{1}\right]\;,\label{eq:FGLS}
\end{equation}
where we note $\hat{R}$ has the same structure as $R$ (\ref{eq:Rform}),
except replacing $\phi$ with $\hat{\phi}$ found from the AR estimator
(\ref{eq:OLS}). With Corollary \ref{cor:phi_bound} we can now bound
the inflation of the errors after the feasible correction, that is,
we consider upper bounding $\|\hat{R}^{\top}\hat{R}L\|_{F}$ and $\|\hat{R}^{\top}\hat{R}L\|_{2}$
and use Lemma \ref{lemma:tail} to choose a $\bar{\lambda}_{n}$ that
ensures $\text{\ensuremath{\mathcal{E}_{\bar{\lambda}}:=\{n^{-1}\|(\hat{R}X)^{\top}\hat{R}Lu\|\le2\bar{\lambda}_{n}\}}}$
occurs in high probability. This choice, alongside the conditions
required for the RE condition to hold on $\hat{R}X$ enable us to
construct an error bound for the FGLS procedure as below.

\begin{proposition}{FGLS Error Bound}\label{prop:fgls}

Let $\bar{\Delta}=\bar{\beta}-\beta$. Assume the conditions and sample
size requirements of Corollary \ref{cor:phi_bound}. Letting the second-stage
regularization parameter be given as
\[
\bar{\lambda}_{n}=\frac{K_{X}K_{u}\|\pi\|_{2}}{c_{1}^{1/2}}a_{n}\delta(n,p^{\tau})
\]
where $a_{n}=1+c_{4}q^{2}v^{2}\delta(n,p^{\tau})$ with positive constant
$c_{4}$, then the FGLS error is bounded according to
\[
\|\bar{\Delta}\|_{2}\le\frac{6K_{X}K_{u}}{c_{1}^{1/2}\kappa}a_{n}\sqrt{\frac{s\log p}{n}}\;,
\]
in probability at least $1-20p^{1-\tau}$. 

\end{proposition}

The result shows that with sufficient samples (scaling with $v^{2}$),
the FGLS estimator with AR($q$) errors can provide a tighter bound
than the corresponding analysis for the regular LASSO. In particular,
we note that $a_{n}\rightarrow1$ as $n$ grows. The error can be
bound to within a constant of that incurred by the LASSO in the oracle
setting where the errors are perfectly corrected, i.e. when $R$ is
known. Given we have assumed a stationary error process, the asymptotic
rate remains the same $\|\bar{\Delta}\|_{2}^{2}=\mathcal{O}_{p}(s\log p/n)$
as the uncorrected LASSO, however, FGLS will maintain better finite
sample performance. The following section investigates this behavior
empirically and provides additional support for the benefits of our
FGLS proposal.

\section{Experimental Results\label{sec:Experimental-Results}}

\subsection{Simulation Setup}

In this Section, we present the Monte Carlo simulation results in
terms of estimation error and sign recovery for the LASSO, GLS-LASSO
and FGLS-LASSO, for a data-generating process as described by the
model (\ref{eq: regression}) with varying degrees of error persistence.
For the simplicity of exposition GLS-LASSO and FGLS-LASSO will be
referred to as GLS and FGLS hereafter. In our setup, we assume a sparsity
parameter of $s=p/10$, such that for $p=100$, the design is simulated
independently across rows according to $X_{t\cdot}\sim N(0,\Sigma)$.
The estimators $\hat{\beta}$ are obtained for GLS and FGLS according
to the description in Section \ref{sec:rgls}. Finally, given the
temporal nature of the regression, the regularization parameter $\lambda_{n}$
is tuned using 2-fold cross-validation, where the folds are defined
as the first and second halves of the data to preserve temporal ordering.

We conduct the simulation study for sample sizes $n=50,100,\cdots,450,500$
and with different numbers of covariates, $p=128,256,512$. We further
consider different autocorrelation parameters $\rho\in\{0,0.5,0.9,0.99\}$
where $\rho=0$ is the classic case of independent errors, where the
LASSO and GLS are expected to perform on par or superior to the FGLS
estimator. On the other hand, with $\rho=0.9$ and $\rho=0.99$, the
error is highly persistent, and realizations would look similar to
those of a random walk. The Monte Carlo study entails $1000$ runs
for each of the aforementioned scenarios, whereby the errors are measured
by $\lVert\hat{\Delta}\rVert_{q}$ for $q=1,2,\infty$, and sign-recovery
is calculated by the empirical probability $\hat{P}[\mathrm{sign}(\hat{\beta})=\mathrm{sign}(\beta_{0})]$,
and $\hat{P}$ is estimated by averaging over the 1000 replications.
Note that $\mathrm{sign}(\hat{\beta})=\mathrm{sign}(\beta_{0})$ requires
equality for all elements, i.e. $\mathrm{sign}(\hat{\beta}_{i})=\mathrm{sign}(\beta_{0;i})$
for all $i=1,\ldots,p$.

\subsection{Results}

\begin{figure}
\begin{centering}
\includegraphics[width=0.6\columnwidth]{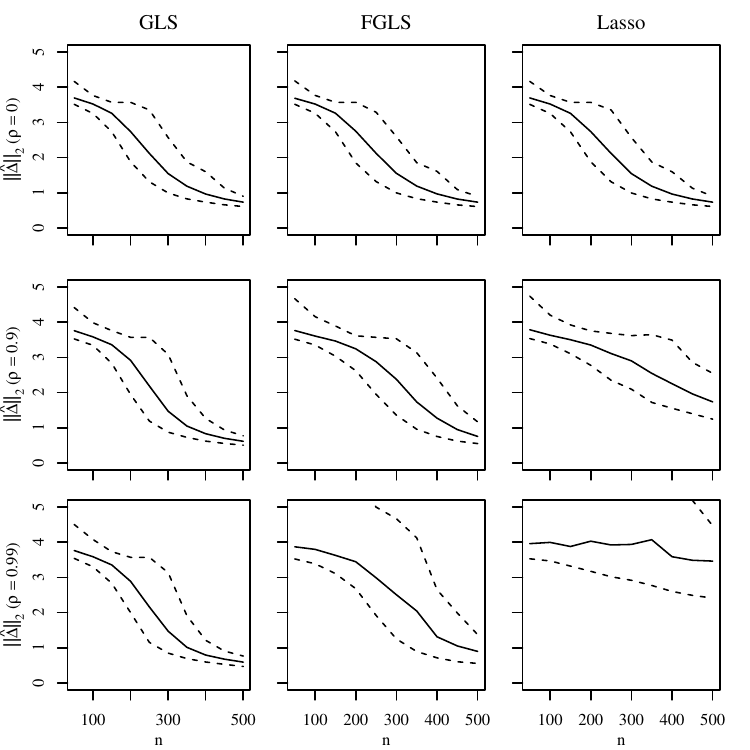}
\par\end{centering}
\begin{centering}
\includegraphics[width=0.6\columnwidth]{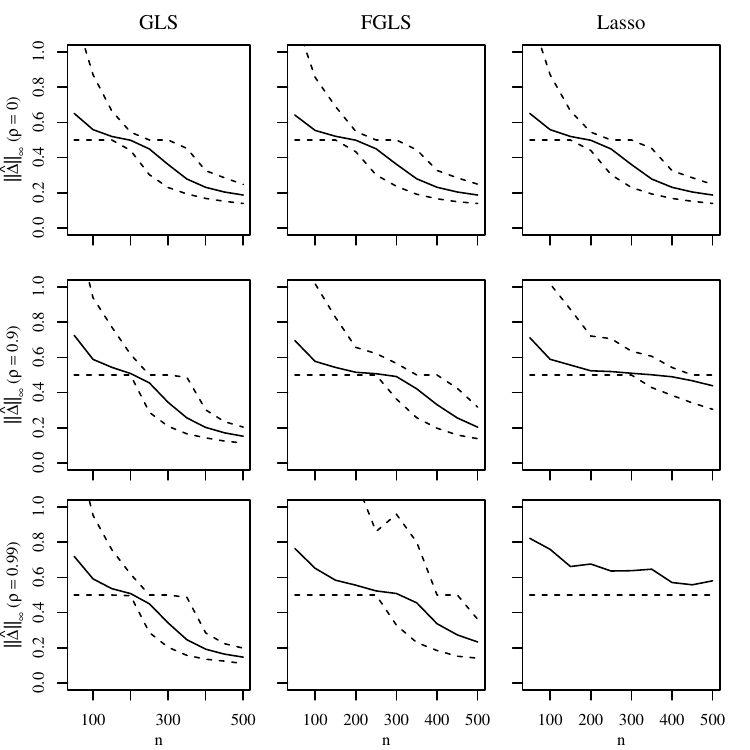}
\par\end{centering}
\caption{Estimation error (top: $p^{-1/2}\|\hat{\Delta}\|_{2}$, bottom: $\|\hat{\Delta}\|_{\infty}$error)
as a function of $n$ for $p=512$ for different settings of $\rho$,
dashed lines indicate empirical 95\% confidence intervals.\label{fig:estimation_error}}
\end{figure}

Figure \ref{fig:estimation_error} present the outcome of the simulations
in terms of the mean $\ell_{2}$ norm of the differences between the
true parameter vector $\beta_{0}$ and the their estimated counterpart
$\hat{\beta}$, using the three approaches of GLS, FGLS and LASSO
for the high-dimensional case with $p=512$. The $95\%$ (empirical)
confindence bands have also been included in the plots to show the
degree of dispersion under different autocorrelation regimes. Similar
results have been compiled using both the $\ell_{1}$and $\ell_{\infty}$-norms,
as well as lower-dimensional cases of $p=128,256$ with similar outcomes.
These results can be found in the supplementary material.

As shown in Fig. \ref{fig:estimation_error}, the performances of
the three estimators are comparable when the errors exhibit zero to
a moderate autocorrelation, i.e., $\rho=0$ and $\rho=0.5$. On the
other hand, the lower-panel with $\rho=0.9$ shows that while the
performance of GLS and FGLS are comparable, they significantly out-perform
the LASSO. As one may expect, given that the FGLS estimator is predicated
on $\hat{\rho}$ as opposed to the true autocorrelation parameter
$\rho$, the FGLS estimates appear to exhibit higher variance. In
the case of the LASSO, the larger degree of dispersion aroud the mean
error is striking, with a 95\% confidence band almost twice that of
both the GLS and FGLS estimators. Moreover, the convergence rate of
LASSO to zero is slower than both of our proposed estimators, with
a mean $\ell_{2}$-error of $\approx1.5$ for $n=500$ for LASSO,
while with an $\ell_{2}$-error of less than $1$ for the GLS and
FGLS estimators. The most interesting results are presented in the
bottom-right panel (Fig. \ref{fig:estimation_error}), where with
$\rho=0.99$, the autocorrelation parameter is close to unity. While
the GLS and FGLS estimators demonstrate similar performances to the
earlier results, LASSO does not show any signs of consistency, with
the $\ell_{2}$-error being significantly larger and approximately
$3$ to $6$ times greater than both the GLS and FGLS estimators.

\begin{figure}
\begin{centering}
\includegraphics[width=0.6\columnwidth]{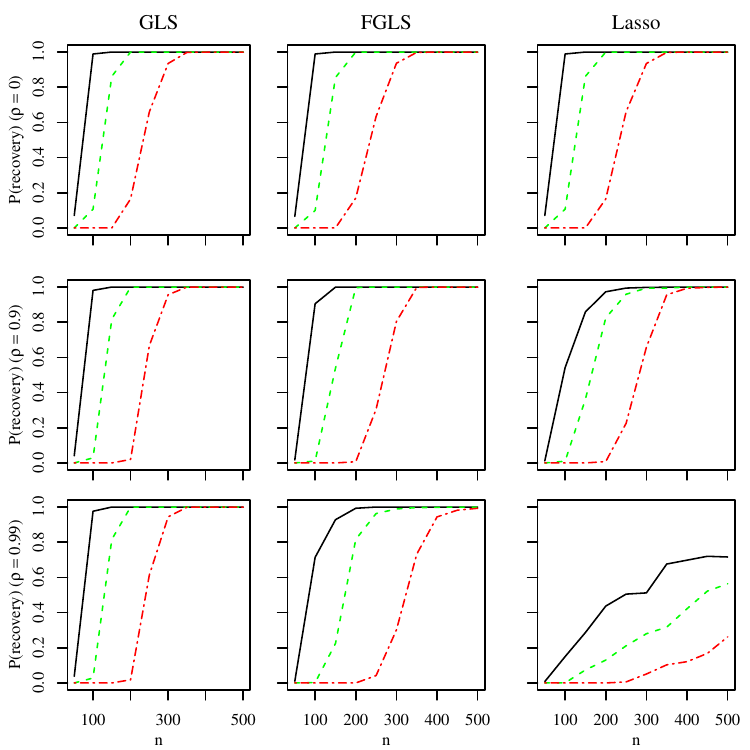}
\par\end{centering}
\caption{Empirical probability of sign recovery $P(\mathrm{sgn}(\hat{\beta})=\mathrm{sgn}(\beta_{0}))$.\label{fig:sign_recovery}}
\end{figure}

Similar findings can be observed in the simulations concerning support
recovery. Sign recovery performances are identical between the different
estimators when the error process exhibits zero or moderate autocorrelation.
However, Figure \ref{fig:sign_recovery} demonstrates a significant
difference between the estimators when the errors are highly persistent.
When $\rho=0.9$ and $p=128,256$ and $512$, although the performances
of the estimators are comparable, the GLS and FGLS clearly outperform
the LASSO. A more interesting case is when $\rho=0.99$, where even
in the low-dimensional settings - i.e., $p=128$, sign recovery for
LASSO and FGLS converge towards $100\%$ with a relatively small sample
size of $n\approx100$, whereas even with $n=500$ the LASSO only
recovers about $70\%$ of the support. This contrast becomes increasingly
more evident in high-dimensional settings, i.e, $p=512$, where the
GLS and FGLS estimators eventually recover the entire support, yet
the LASSO only recovers roughly $25\%$ of the support when $n=500$.

\section{Discussion\label{sec:Conclusion}}

This paper provides a detailed study on the properties of a simple
regularized GLS procedure for linear regressions with potentially
autocorrelated error terms, in a high-dimensional setting. We study
the case with both sub-Gaussian design and errors, where the error
process is assumed to be autoregressive in nature. Our contributions
are three-fold: 

First, we confirm that in the presence of autocorrelated errors and
without the GLS (or FGLS) transformation, the choice of the regularization
parameter should be inflated in relation to the degree of persistency
of the error terms. A quick glance at Corollary \ref{cor:ar_deviation_bound}
reveals that for any fixed $p,n$, we have $\lambda\propto(1-\rho^{2})^{-1}$,
hence $\lambda\to\infty$ if $|\rho|\to1$. Consequently, the LASSO's
error rate (under the $\ell_{2}$ norm) can slow as the autocorrelation
of the error process approaches unity. These theoretical results are
further fortified by the Monte Carlo simulations exercise in Section
4, specifically where we study the highly persistent error setting
e.g. $\rho=0.99$. 

Second, in the case of the GLS estimator, we show a restricted eigenvalue
condition for the transformed design matrix holds in high-probability.
This result generalizes those found in \cite{Yu2010,Rudelson2012}
to the autocorrelated setting, where our whitening matrix induces
such autocorrelation. Once this is obtained, we subsequently provide
an oracle-inquality for the GLS-LASSO, which holds over a range of
supports of bounded size. Crucially, we demonstrate that when whitening
is performed in relation to the AR errors, the eigenvalues of the
design are still adequately lower-bounded. We note that this is in
contrast to the re-scaling that the design may undergo with different
error assumptions, e.g., as in \cite{Jia2015}---whilst we have shown
GLS-LASSO works well with autoregressive errors, it may not generally
be optimal to perform the GLS rotation.

Third, we present non-asymptotic bounds for the parameter in the AR(1)
errors that take into account estimation error (in the regression
coefficients) from the first-stage of the FGLS-LASSO. While the asymptotic
consistency of this parameter using FGLS is well-established in classical
settings ($p<n$), to our knowledge, our work is the first of its
kind in the high-dimensional setting, using the LASSO. We further
use this result to enable the construction of an error bound for the
FGLS procedure through an appropriate choice of regularizer parameter
in the second stage estimation. Our result shows that whilst the choice
of $\bar{\lambda}_{n}$ is inflated slightly compared to the nominal
choice (in presence of iid errors) this inflation is bounded, and
asymptotically $n,p$ tends to zero for $n=\Omega(q^{4}s^{2}\log p)$.

Finally, our simulation exercises in Section 4, compare the LASSO,
GLS-LASSO and FGLS LASSO estimators in terms of estimation error and
sign-recovery for different levels of persistence in the error terms.
Our results in this section corroborate the theoretical findings obtained
in terms of estimation error, and show that FGLS can perform well
in both high- and low-dimensional settings. Furthermore, while we
have not presented theoretical results pertaining to sparsistency,
the simulations indicate the superiority of the GLS-LASSO and FGLS-LASSO
in terms of sign recovery when the errors are highly persistent, whilst
giving identical performance in the absence of the correlation of
the error terms. Theoretical results for sparsitency could be derived
by extending the now standard primal-dual witness argument, c.f..
\cite{wainwright2019high}, where one would need to consider the impact
of the whitening matrix on incoherence type conditions.

Our work further paves the path for future research for performing
GLS type corrections within a regularized M-estimation framework.
Beyond extension to other structured penalties/priors, c.f., the group-LASSO,
trend-filtering, we can also consider relaxing the distributional
assumptions used in our results, for example looking at settings where
the autocorrelation structures may vary over time, or where the error
structure can be approximated by an AR($q_{n})$ process. To some
extent, the work of \cite{chronopoulos2023high} provides work in
this direction, where they allow for a wider range of error dependencies
and also consider the task of constructing valid confidence intervals
for regression parameters (c.f. \cite{VanDeGeer2014}), however, there
are still key differences between that work and our own, in particular
with regards to the analysis of the RE condition in the FGLS setting
where we focus on understanding the impact of the AR($q$) whitening
matrix on the original design matrix. Overall, we have shown that
GLS (and a feasible variant) can be highly effective when combined
with the LASSO.

\section{Appendix}

\subsection{Proofs for Section 2. and Preliminaries}

\begin{proof}{Lemma \ref{lemma:tail}}

Consider that $\|X^{\top}Lu\|_{\infty}=\max_{j}|w_{j}|$ where the
$j$ specific variable is given by
\[
w_{j}=\sum_{t=1}^{n}x_{tj}\sum_{k=1}^{n}L_{tk}u_{k}\;.
\]
Hence we desire a deviation bound on $w_{j}$ for each $j\in[p]$.
Note that $e_{t}=\sum_{k=1}^{n}L_{tk}u_{k}$ is a sub-Gaussian random
variable with norm $K_{e}\le K_{u}\|L_{t,\cdot}\|_{2}$, thus we can
write $w_{j}$ as a sum of the product of sub-Gaussian random variables.
Our strategy will be to bound the moment generating function (MGF)
of $w_{j}$ and then apply a Bernstein type argument.

For notational convenience let $x:=X_{\cdot j}$ for the $j$th column
of $X$, such that we can then write $w_{j}=x^{\top}Lu$, with $u=(u_{1},\ldots,u_{n})$.
We note the conditional distributions $w_{j}|u$ and $w_{j}|x$ are
a sub-Gaussian (zero-mean) random vector with norm bounded $K_{X}\|Lu\|_{2}$
or $K_{u}\|x^{\top}L\|_{2}$ respectively. Now consider $w_{j}|u$,
for which we can obtain the bound on the MGF as
\begin{equation}
\mathbb{E}_{x|u}[\exp(\eta x^{\top}Lu)]\le\exp(c_{1}\eta^{2}K_{X}^{2}\|Lu\|_{2}^{2})\;,\quad\forall\eta\in\mathbb{R}\;,\label{eq:conditional_mgf}
\end{equation}
using the definition of $\normiii{w_{j}|u}_{2}$. Compare this to
the bound obtained if $x$ were replaced with the Gaussian random
variable $z$, where we would obtain
\begin{equation}
\mathbb{E}_{z|u}[\exp(az^{\top}Lu)]=\exp(a^{2}\|Lu\|_{2}^{2}/2)\;,\label{eq:gaussian_mgf}
\end{equation}
again for any constant $a$. We can choose $a=\sqrt{2c_{1}}K_{X}\eta$,
and now match the right hand sides of (\ref{eq:conditional_mgf})
and (\ref{eq:gaussian_mgf}), thus we obtain
\[
\mathbb{E}_{x|u}[\exp(\eta x^{\top}Lu)]\le\mathbb{E}_{z|u}[\exp(\sqrt{2c_{1}}K_{X}\eta z^{\top}Lu)]\;.
\]
We can therefore effectively replace $x$ with the Gaussian variable
$z$. To find an overall upper bound we compute the expectation
\[
\mathbb{E}_{u}[\exp(c_{1}\eta^{2}K_{X}^{2}\|Lu\|_{2}^{2})]=\mathbb{E}_{u}[\exp(c_{1}\eta^{2}K_{X}^{2}u^{\top}L^{\top}Lu)]\;.
\]
Since $u$ has independent mean-zero sub-Gaussian coordinates with
norm at most $K_{u}$, the quadratic form $u^{\top}L^{\top}Lu$ satisfies
a Hanson-{}-Wright type MGF bound, c.f. Proof of Theorem 1.1 in \cite{Rudelson2013};
in particular, there exist absolute constants $c_{2},c_{3}>0$ such
that
\[
\mathbb{E}_{u}\exp\Big(bu^{\top}L^{\top}Lu\Big)\le\exp\Big(c_{2}bK_{u}^{2}\|L\|_{F}^{2}\Big)\qquad\text{for }0\le b\le c_{3}/K_{u}^{2}\|L\|_{2}^{2}.
\]
Now substitute $b=c_{1}\eta^{2}K_{X}^{2}$, to find 
\[
\mathbb{E}[\exp(\eta w_{j})]\le\exp\big(c_{4}\eta^{2}K_{X}^{2}K_{u}^{2}\|L\|_{F}^{2}\big)\quad\text{for }0\le\eta\le c_{5}/K_{u}K_{X}\|L\|_{2}\;.
\]
To complete the proof, we apply the Markov inequality and optimize
over the exponent:
\begin{align*}
P[\exp(\eta w_{j})\ge\exp(\eta\epsilon)] & \le\frac{\mathbb{E}[\exp(\eta w_{j})]}{\exp(\eta\epsilon)}\\
 & \le\exp(-\eta\epsilon+c_{4}K_{X}^{2}K_{u}^{2}\eta^{2}\|L\|_{F}^{2})
\end{align*}
considering the range $0\le\eta\le c_{5}/K_{X}K_{u}\|L\|_{2}$. Optimizing
over the exponent leads to a choice of $\eta^{*}=\min[\epsilon\big/2c_{4}K_{X}^{2}K_{u}^{2}\|L\|_{F}^{2},c_{5}/K_{X}K_{u}\|L\|_{2}]$,
leading to
\[
P[|w_{j}|\ge\epsilon]\le2\exp\left[-c\min\bigg(\frac{\epsilon^{2}}{K_{X}^{2}K_{u}^{2}\|L\|_{F}^{2}},\frac{\epsilon}{K_{X}K_{u}\|L\|_{2}}\bigg)\right]\;,
\]
for some $c>0$. Applying the union bound over $j\in[p]$ completes
the proof.

\end{proof}

\begin{lemma}{Hanson-Wright Inequality}\label{lemma:hanson_wright}

Let $u=(u_{1},\ldots,u_{n})^{\top}$ be a random sub-Gaussian isotropic
vector with $\max_{t}\normiii{u_{t}}_{2}\le K_{u}$. Given a deterministic
matrix $A\in\mathbb{R}^{n\times n}$, we have the bound
\begin{equation}
P\left[|u^{\top}Au-\mathrm{tr}(A)|\ge\epsilon\right]\le2\exp\left[-c\min\left(\frac{\epsilon^{2}}{K_{u}^{4}\lVert A\rVert_{F}^{2}},\frac{\epsilon}{K_{u}^{2}\lVert A\rVert_{2}}\right)\right]\;,\label{eq:chaos_HW}
\end{equation}
where $c>0$ is an absolute constant.

\end{lemma}

\begin{proof}{Hanson-Wright - Lemma \ref{lemma:hanson_wright}}See
for instance \cite{Rudelson2013}, or Chapter 6 in \cite{vershynin2018high}.
\end{proof}

\begin{proof}{Corollary 1}

Recall, we want to bound the probability of the event $\mathcal{E}_{\lambda}:=\{2n^{-1}\|X^{\top}\Psi_{q}u\|_{\infty}<\lambda_{n}\}$.
Note that $\|\Psi_{q}\|_{F}^{2}=nv^{2}$ and $\|\Psi_{q}\|_{2}=v$,
from Lemma \ref{lemma:tail}, we have

\begin{align*}
P[n^{-1}\|X^{\top}e\|_{\infty}\ge\epsilon/2] & \le2p\exp\left(-c\min\left[\frac{\epsilon^{2}n}{4K_{X}^{2}K_{u}^{2}v^{2}},\frac{\epsilon n}{2K_{X}K_{u}v}\right]\right)\\
 & =2p\exp\left(-\frac{c\epsilon n}{2K_{X}K_{u}v}\min\left[\frac{\epsilon}{2K_{X}K_{u}v},1\right]\right)
\end{align*}
Now choose $\lambda_{n}=2c^{-1/2}K_{X}K_{u}v\sqrt{\log p^{\tau}/n}$
giving

\begin{align*}
P[\mathcal{E}_{\lambda}^{c}] & \le2p\exp\left(-cn\frac{\lambda_{n}^{2}}{4K_{X}^{2}K_{u}^{2}v^{2}}\right)=2p\exp(-\delta^{2}\left(1,p^{\tau}\right))\;.\\
 & =2p^{1-\tau}
\end{align*}
where we have used $n\ge c^{-1}\tau\log p$ which forces the choice
of the Gaussian tail.

\end{proof}

\begin{proof}{Lemma \ref{lemma:ma_comparison}}

For most entries in $R^{\top}R$, one can consider the matrix to be
equivalent to $\Gamma_{\mathrm{MA}(\pi)}$. The only differences are
in the first $i,j<q$ rows/columns, and the last $i,j>n-q$ rows/columns.
On inspection, one can see these \emph{initialization} blocks take
the form

\[
R^{\top}R=\left(\begin{array}{ccccc}
{\color{blue}a_{1}} & \mathbin{\color{blue}-}{\color{blue}\phi_{1}} & {\color{blue}} & \mathbin{\color{blue}-}{\color{blue}\phi_{q-1}} & \gamma_{q}\\
\mathbin{\color{blue}-}{\color{blue}\phi_{1}} & {\color{blue}\ddots} & \mathbin{\color{blue}-}{\color{blue}\phi_{1}} & {\color{blue}} & \gamma_{2}\\
{\color{blue}} & {\color{blue}} & {\color{blue}} & \mathbin{\color{blue}-}{\color{blue}\phi_{1}}\\
\mathbin{\color{blue}-}{\color{blue}\phi_{q-1}} & {\color{blue}} & \mathbin{\color{blue}-}{\color{blue}\phi_{1}} & {\color{blue}a_{q-1}} & \gamma_{1}\\
\gamma_{q} & \cdots &  & \gamma_{1} & a_{q}
\end{array}\right)
\]
with $a_{1}=1$ and $a_{k}=a_{k-1}+\phi_{k-1}^{2}$ integrates the
variance of the MA($q$) model until the stationary variance is obtained,
i.e. when $i,j\ge q$. Note: the discrepancy from $\Sigma_{\mathrm{MA}(\pi)}$
is highlighted by the blue entries above, and for $i,k\ge q$ we obtain
the stationary Toeplitz form. Due to symmetry such a block also occurs
at the end of the matrix. 

For the maximum eigenvalue, we can apply Gershogorin's circle theorem,
let 
\[
r_{i}=\sum_{j\ne i}|[R^{\top}R]_{ij}|
\]
and $\mathcal{D}([R^{\top}R]_{ii},r_{i})\subseteq\mathbb{C}$ be a
closed disc centered at $[R^{\top}R]_{ii}$ . We know every eigenvalue
of $R^{\top}R$ lies in one of these discs. We bound the size of each
disk claiming that the largest magnitude (within disc) for $\Sigma_{MA(\pi)}$
is greater than that for $R^{\top}R$. To show this we need to show
that the maximum row sum of $\Sigma_{MA(\pi)}$ dominates $R^{\top}R$
for all rows. 

The largest row for $R^{\top}R$ that does not coincide with $\Sigma_{MA(\pi)}$
is given by the vector
\[
r_{1}:=(-\phi_{q-1},\ldots,-\phi_{1},a_{q},\gamma_{1},\ldots,\gamma_{q})
\]
which we compare with
\[
r_{2}:=(\gamma_{q},\gamma_{q-1},\ldots,\gamma_{1},\gamma_{0},\gamma_{1}\ldots,\gamma_{q})\;.
\]
If $\|r_{2}\|_{1}\ge\|r_{1}\|_{1}$ then we can use $\|\Sigma_{MA(\pi)}\|_{2}$
as an upper bound for $\|R^{\top}R\|_{2}$. We thus need to check
\begin{align*}
\sum_{i=0}^{q}|\gamma_{i}| & \ge\sum_{i=1}^{q-1}|\phi_{i}|+a_{q}
\end{align*}
which is implied if $\sum_{i=1}^{q}|\gamma_{i}|-\phi_{q}^{2}-\sum_{i=1}^{q-1}|\phi_{i}|\ge0$.
Recall $\gamma_{i}=\sum_{k=0}^{q-i}\pi_{k}\pi_{k+i}=-\phi_{i}+\sum_{k=1}^{q-i}\phi_{k}\phi_{k+i}$
then in general it would be sufficient to show
\[
\sum_{i=1}^{q}\bigg|-\phi_{i}+\sum_{k=1}^{q-i}\phi_{k}\phi_{k+i}\bigg|-\phi_{q}^{2}-\sum_{i=1}^{q-1}|\phi_{i}|\ge0
\]
We will show the bound holds explicitly for AR(1) and AR(2), and leave
this as a conjecture for higher order models.
\begin{itemize}
\item AR(1) requires $|\phi_{1}|-\phi_{1}^{2}\ge0$ which is implied if
$|\phi_{1}|(1-|\phi_{1}|)\ge0$, and is thus true for all $|\phi|<1$
in the stationary regime.
\item AR(2) requires:\emph{
\begin{align*}
f(\phi_{1},\phi_{2}):= & |-\phi_{1}+\phi_{1}\phi_{2}|+|\phi_{2}|-\phi_{2}^{2}-|\phi_{1}|\\
= & |\phi_{1}(\phi_{2}-1)|-|\phi_{1}|+|\phi_{2}|(1-|\phi_{2}|)\ge0\;.
\end{align*}
}We note this is symmetric in $\phi_{1}$, and $f(\phi_{1},0)=0$,
for $\phi_{2}\le0$ we see $f()$ is monotonically increasing as a
function of $\phi_{1}$ and $f(0,\phi_{2})=|\phi_{2}|(1-|\phi_{2}|)\ge0$
for all $|\phi_{2}|\le1$. Finally, we check the upper quadrant $\phi_{1}\ge0,\phi_{2}\ge0$
for which $f()$ is monotonically decreasing as a function of $\phi_{1}$,
and we have already shown that $f(0,\phi_{2})\ge0$. We thus need
to check the border of the stationary region i.e. the line $\phi_{2}=1-\phi_{1}$
for which we find
\[
|\phi_{1}(\phi_{2}-1)|-|\phi_{1}|+|\phi_{2}|(1-|\phi_{2}|)=|\phi_{1}^{2}|-|\phi_{1}|+|1-\phi_{1}|(1-|1-\phi_{1}|)
\]
which is positive for all $\phi_{1}\in[0,1]$.
\end{itemize}
Applying Gershogorin's circle theorem on the row $r_{2}$ to find
$\|R^{\top}R\|_{2}\le\|r_{2}\|_{1}\le(2q+1)\|\pi\|_{2}^{2}$ which
completes the result.

\end{proof}

\subsection{Proof of Proposition X (sub-Gaussian)\label{subsec:subGaussian_proof}}

\begin{proof}{Proposition \ref{prop:re_subgaussian} - Sub-Gaussian RE}

The action of $R$ on $X$ will result in a larger variance for the
rotated design as $\|\pi\|_{2}^{2}\ge1$, for convenience, in the
following arguments we factor this out and study the decomposition
\[
\tilde{X}=R\bar{W}\bar{\Sigma}^{1/2}
\]
where $\bar{W}=W/\|\pi\|_{2}$ and $\bar{\Sigma}=\|\pi\|_{2}\Sigma^{1/2}$.
Our approach now is to examine the concentration of $R\bar{W}$ when
projected onto the unit sphere, this allows us to examine how the
random matrix will impact the scale of a vector on which it acts.
We then use this concentration result applied to a vector $\bar{\Sigma}v$
to examine the impact of rotation on vectors $v\in\mathbb{C}_{\alpha}(\mathcal{S})$
and subsequently show that a restricted eigenvalue condition on the
rotated design will hold in high probability. The following Lemma
forms the first part of our argument, applying the Hanson-Wright inequality
(Lemma \ref{lemma:hanson_wright}) in conjunction with our bounds
on the change in variance due to $R$.

\begin{lemma}{Design Deviation Bound}\label{lemma:deviation_design}

Assuming $R=\Psi_{q}^{-1}$, then for all $z\in\mathbb{S}^{p-1}:=\{z\in\mathbb{R}^{p}\:|\:\|z\|_{2}=1\}$
, we have
\[
1-q/n\le\mathbb{E}[n^{-1}\|R\bar{W}z\|_{2}^{2}]\le1
\]
and for $q/n<\epsilon<K_{W}^{2}\|\pi\|_{2}(q+1)$ we find
\[
P[|\frac{1}{n}\|R\bar{W}z\|_{2}^{2}-1|>\epsilon]\le2\exp\left(-c\frac{\epsilon^{2}n}{K_{W}^{4}\|\pi\|_{2}^{2}(q+1)^{2}}\right)\;,
\]
for some absolute constant $c>0$.

\end{lemma}

In the above, (see Section \ref{prop:re_subgaussian} for proof) we
state the result choosing a Gaussian tail corresponding to quadratic
decay in the tail in terms of $\epsilon$, which occurs if $\epsilon$
is sufficiently small. However, if we are interested in larger deviations,
or alternative designs one could equally study the exponential tail.
We see that as $n\rightarrow\infty$ the concentration of $n^{-1}\|R\bar{W}z\|_{2}^{2}$
is quickly towards the expectation of one, however, in finite samples
there can still be some deviation due to the initialization blocks
in $R$, which can be seen as an edge-effect. Our next step is to
show the scale of a vector $Av$ (we will later use $A=\bar{\Sigma}$)
is not magnified too much when multiplied by $R\bar{W}$.

\begin{lemma}{Almost Isometry Property (Post Rotation)}\label{lemma:UUP_post_rot}

Let $A\in\mathbb{R}^{p\times p}$, $R=\Psi_{q}^{-1}$, and consider
an $r$-sparse vector $v\in\mathbb{R}^{p}$. Under Assumption \ref{ass:1}
and with sufficient samples
\begin{equation}
n\ge c^{-1}K_{W}^{4}\|\pi\|_{2}^{2}(q+1)^{2}\left[\frac{\log(p^{\tau})+r\log(6p\big/r\epsilon)}{\epsilon^{2}}\right]\label{eq:min_sample}
\end{equation}
for $n^{-1}q<\epsilon\le\min[2^{-1},K_{W}^{2}\|\pi\|_{2}(q+1)]$ we
have
\begin{equation}
(1-2\epsilon)\|Av\|_{2}\le n^{-1/2}\|R\bar{W}Av\|_{2}\le(1+2\epsilon)\|Av\|_{2}\;,\label{eq:UUP}
\end{equation}
in probability greater than $1-2p^{-\tau}$.

\end{lemma}

The result is a modification of a bound from Rudleson et al. \cite{Rudelson2012}
where we combine Lemma \ref{lemma:deviation_design} with an $\epsilon$-net
argument to control the size of the projections $R\bar{W}z$ on the
sphere, see Section \ref{subsec:subGaussian_proof} for a detailed
proof. This result holds for a range of different levels of sparsity
$r$, however, we see the minimum sample size should grow with $r$.
Additionally, for small $\epsilon$ we should still need a large sample
size, note that the term $\log(6\mathrm{e}p\big/r\epsilon)$ will
be significant for all $r\epsilon<1$, however, we would otherwise
have a requirement for $n=\Omega(q^{2}r\log p^{\tau})$ observations.
Importantly, the impact of the whitening matrix is minimal, where
we can compare to the independent error situation by setting $\|\pi\|_{2}^{2}(q+1)^{2}=1$.

Whilst we state the result for the AR($q$) case here, the argument
is extensible to a generic $R$ in which case one needs to consider
how $n^{-1}\|R^{\top}R\|_{F}^{2}$ scales in place of $q\|\pi\|_{2}^{2}$.
In our case, the difference in structure for the first $q$ rows (i.e.
the impact of $R_{I}$) is quickly negligible, e.g., under the required
sample size condition. We now apply the above result to cover the
set of sparse vectors in the cone $\mathbb{C}_{\alpha}(\mathcal{S})$
using a contraction result from \cite{Rudelson2012}.

The proof is now a consequence of a reduction principle proved in
\cite{Rudelson2012} and stated below, applied in conjunction with
Lemma \ref{lemma:UUP_post_rot}. 

\begin{lemma}{Theorem 3 - Ruddelson 2012 \label{lemma:contraction}}

Assume that for $A\in\mathbb{R}^{p\times p}$ and some $\kappa>0$,
we have
\[
\|Az\|_{2}\ge\kappa\|z\|_{2}\quad\forall z\in\mathbb{C}_{3\alpha}(\mathcal{S})
\]
for all subsets $\mathcal{S}$ such that $|\mathcal{S}|=s<p$. Now
for $1>\zeta>0$, consider the effective sparsity
\begin{equation}
d=s\left(1+\frac{5^{2}2^{4}3^{2}\alpha^{2}(3\alpha+1)}{\kappa^{2}\zeta^{2}}\max_{j\in\mathcal{S}}\|\Sigma^{1/2}b_{j}\|_{2}^{2}\right)\;,\label{eq:s_expanded}
\end{equation}
where $b_{j}$ is a canonical basis vector for $j\in\mathcal{S}$.
Further consider $\mathcal{B}_{d}=\cup_{|\mathcal{S}|=d}\mathcal{B}_{\mathcal{S}}\subseteq\mathbb{R}^{p}$
is the subspace spanned by $\{b_{j}\}_{j\in\mathcal{S}}$ if $d<p$,
and $\mathcal{B}=\mathbb{R}^{p}$ otherwise. For a (deterministic)
matrix $G\in\mathbb{R}^{n\times p}$ which obeys 
\begin{equation}
(1-\zeta/5)\|u\|_{2}\le\|Gu\|_{2}\le(1+\zeta/5)\|u\|_{2}\quad\forall\:u=Av\;,v\in\mathcal{B}_{d}\;.\label{eq:uup_contraction}
\end{equation}
then
\[
\|GAz\|_{2}\ge\kappa'\|z\|_{2}\quad\forall z\in\mathbb{C}_{\alpha}(\mathcal{S})
\]
for some $0<(1-\zeta)\kappa\le\kappa'<\infty\;.$

\end{lemma}

We will apply Lemma \ref{lemma:contraction} and use the representation
$RX=GA$ with $G=n^{-1/2}R\bar{W}$ and $A=\Sigma^{1/2}\|\pi\|_{2}$.
We now apply Lemma \ref{lemma:UUP_post_rot} with $r=\min(d,p)$ defined
in (\ref{eq:s_expanded}), $A=I_{p}$ and $2\epsilon=\zeta/5$. Dividing
(\ref{eq:UUP}) by $\|u\|_{2}$ we find
\[
(1-\zeta/5)\|u\|_{2}\le\|Gu\|_{2}\le(1+\zeta/5)\|u\|_{2}\;,
\]
thus verifying the condition (\ref{eq:uup_contraction}) on $G$,
where $u$ is an $r$-sparse vector. This holds under the probabilities
and sample size conditions previously stated, where we require
\[
n^{-1}q<\zeta<\min[1,10K_{W}^{2}\|\pi\|_{2}(q+1)]\;.
\]
Note, we state the result assuming $\mathcal{E}_{\mathrm{RE}}(\Sigma^{1/2};\kappa,3\alpha)$
holds, rather than $\mathcal{E}_{\mathrm{RE}}(A;\kappa,3\alpha)$
as required by the contraction result, however, given our form of
$A$ this simply results in a rescaling of the eigenvalue by $\|\pi\|_{2}$.

For convenience, we work with an upper bound on the effective sparsity
$d'=Cs\big/\kappa^{2}\zeta^{2}\ge d$, where $C=2^{5}5^{2}3^{2}\alpha^{2}(3\alpha+1)\max_{j\in\mathcal{S}}\|\Sigma^{1/2}b_{j}\|_{2}^{2}$
and simplifying the sample size requirement
\begin{align*}
n & \ge C^{-1}\left[\frac{\log(p^{\tau})+d'\log(\frac{6\mathrm{e}p}{d'\epsilon})}{\epsilon^{2}}\right]\\
\impliedby n & \ge\frac{K_{W}^{4}\|\pi\|_{2}^{2}(q+1)^{2}\alpha^{3}}{c}\max_{j\in\mathcal{S}}\|\Sigma^{1/2}b_{j}\|_{2}^{2}\frac{s}{\kappa^{2}\zeta^{4}}\log(p^{\tau})\;,
\end{align*}
where again $c$ is modified to account for some numerical constants
and we can focus on the $\log(p^{\tau})$ term as long as $p^{\tau-1}\ge6e.$
Applying the contraction result of Lemma \ref{lemma:contraction}
we require $0<\zeta\le1$ and thus obtain $\tilde{\mathcal{E}}_{\mathrm{RE}}$
in the same probability.

\end{proof}

\subsection{Proof of Lemma \ref{lemma:raskutti} }

\begin{proof}{Lemma \ref{lemma:raskutti} - Gaussian RE (Raskutti et al.)}

We wish to bound $\inf_{v\in\mathbb{R}^{p}}\|RXv\|=\inf_{v\in\mathbb{R}^{p}}\|R\bar{W}\bar{\Sigma}^{1/2}v\|_{2}$
where $\bar{\Sigma}^{1/2}=\|\pi\|_{2}\Sigma^{1/2}$ and $\bar{W}=W/\|\pi\|_{2}$
is a scaled isotropic Gaussian random matrix. We note that the argument
here follows very closely that of \cite{Yu2010} except we account
for the dependence in $RX$ across rows. We reproduce the proof here
for completeness, but recommend readers the original paper if they
desire further details.

Without loss of generality we study the variable $v'=v/\|\bar{\Sigma}^{1/2}v\|_{2}$
such that $\|\bar{\Sigma}^{1/2}v'\|_{2}=1$, specifically consider
\[
M(r,W):=\sup_{v\in\mathcal{V}(r)}\left[1-\frac{\|R\bar{W}v\|_{2}}{\sqrt{n}}\right]
\]
where $\mathcal{V}(r):=\{v\in\mathbb{R}^{p}\:|\:\|\bar{\Sigma}^{1/2}v\|_{2}=1,\|v\|_{1}\le r\}$
which can be seen as the intersection of an ellipsoid and the $\ell_{1}$
ball of radius $r$. A key difference in this case (compared to Prop
\ref{lemma:UUP_post_rot}), is that we have a known distribution for
$W$, and we incorporate the $\ell_{1}$ constraint directly into
$\mathcal{V}(r)$ which allows for a direct measure of how the minimum
eigenvalue can depend on $\|v\|_{1}$. We now construct the random
variable
\begin{align*}
Y_{u,v} & =u^{\top}R\bar{W}v\equiv u^{\top}RXv\;,\\
Z_{u,v} & =\frac{1}{\|\pi\|_{2}}g^{\top}R^{\top}u+h^{\top}v\\
 & =\tilde{g}^{\top}u+h^{\top}v
\end{align*}
where $g\sim\mathcal{N}_{n}(0,I_{n})$, $h\sim\mathcal{N}_{p}(0,I_{p})$
and $u\in\mathbb{S}^{n-1}$. The second formulation of $Z_{u,v}$
relies on $\tilde{g}=\|\pi\|_{2}^{-1}Rg\sim\mathcal{N}_{n}(0,\|\pi\|_{2}^{-2}R^{\top}R)$.
Computing the variances across index points $u,v$ and $u',v'$, we
find
\[
\mathrm{Var}[Z_{u,v}-Z_{u',v'}]=\|v-v'\|_{2}^{2}+\frac{1}{\|\pi\|_{2}^{2}}\|R(u-u')\|_{2}^{2}
\]
and
\[
\mathrm{Var}[Y_{u,v}-Y_{u',v'}]=\frac{1}{\|\pi\|_{2}^{2}}\|R\{uv^{\top}-u'(v')^{\top}\}\|_{F}^{2}\;,
\]
furthermore, through the Cauchy-Schwarz inequality we find $\mathrm{Var}[Y_{u,v}-Y_{u',v'}]\le\mathrm{Var}[Z_{u,v}-Z_{u',v'}]$
$\forall(u,v)\in\mathbb{S}^{n-1}\times\mathcal{V}(r)$. We can now
apply Gordon's inequality to find
\begin{align*}
\mathbb{E}[\sup_{v\in\mathcal{V}(r)}\inf_{u\in\mathbb{S}^{n-1}}u^{\top}RXv] & \le\mathbb{E}[\sup_{v\in\mathcal{V}(r)}\inf_{u\in\mathbb{S}^{n-1}}Z_{u,v}]\\
 & =-\mathbb{E}[\|\tilde{g}\|_{2}]+\mathbb{E}[\sup_{v\in\mathcal{V}(r)}h^{\top}\bar{\Sigma}^{1/2}v]
\end{align*}
We further have $\sup_{v\in\mathcal{V}(r)}h^{\top}\bar{\Sigma}^{1/2}v\le r\|\bar{\Sigma}^{1/2}h\|_{\infty}$,
and since we are working with Gaussian random-variables $\mathbb{E}[\|\bar{\Sigma}^{1/2}h\|_{\infty}]\le3\sqrt{\max_{ii}(\bar{\Sigma}_{ii})\log p}$.
Thus 
\[
\mathbb{E}[\sup_{v\in\mathcal{V}(r)}h^{\top}\bar{\Sigma}^{1/2}v]\le3r\|\pi\|_{2}\|\Sigma\|_{\infty}^{1/2}\sqrt{\log p}\;.
\]
For the term $\|\tilde{g}\|_{2}$, we note that this is normalized
such that each entry will have variance less than or equal to one.
Note that $\|\tilde{g}\|_{2}^{2}$ is a generalised $\chi^{2}$ distribution
with 
\[
\mathrm{Var}[\|\tilde{g}\|_{2}^{2}]=\frac{2\tr(RR^{\top}RR^{\top})}{\|\pi\|_{2}^{4}}
\]
\[
\mathbb{E}[\|\tilde{g}\|_{2}^{2}]=\frac{\mathrm{tr}(R^{\top}R)}{\|\pi\|_{2}^{2}}
\]
We use the bound, for random variable $x$, that
\[
\frac{\mathbb{E}[x^{2}]+x^{2}}{2n}-\frac{\mathrm{Var}[x^{2}]}{2n^{2}}\le\frac{x}{n^{1/2}}\le\frac{\mathbb{E}[x^{2}]+x^{2}}{2n}\;,
\]
to bound
\begin{align*}
\frac{1}{n^{1/2}}\mathbb{E}[\|\tilde{g}\|_{2}] & \ge\frac{1}{n}\frac{\|R\|_{F}^{2}}{\|\pi\|_{2}^{2}}-\frac{1}{n^{2}}\frac{\|R^{\top}R\|_{F}^{2}}{\|\pi\|_{2}^{4}}\\
 & \ge\frac{n-q}{n}-\frac{(2q+1)^{2}}{n}\\
 & \ge1-\frac{1}{n}[q+2(q+1)^{2}]\\
\implies\mathbb{E}[\|\tilde{g}\|_{2}] & \ge(1-\omega)\sqrt{n}
\end{align*}
where on the second line we have used $\|R\|_{F}^{2}\ge(n-q)\|\pi\|_{2}^{2}$
and $\|R^{\top}R\|_{F}^{2}\le(2q+1)^{2}\|\pi\|_{2}^{2}n$, and the
last line requires $\omega$ set to be sufficiently large $\omega=\Omega(n^{-1/2}q^{2})$. 

Recalling the definition of $M(r,W)$ we see
\begin{align*}
\mathbb{E}\left[-\inf_{v\in\mathcal{V}(r)}\|RXv\|_{2}\right] & \le-\sqrt{n}(1-\omega)+3r\|\pi\|_{2}\|\Sigma\|_{\infty}^{1/2}\sqrt{\log p}\\
\mathbb{E}[M(r,W)]=\mathbb{E}\bigg[1-\inf_{v\in\mathcal{V}(r)}n^{-1/2}\|R\bar{W}v\|_{2}\bigg] & \le\underset{:=t(r)}{\underbrace{\omega+3r\|\pi\|_{2}\|\Sigma\|_{\infty}^{1/2}\sqrt{\frac{\log p}{n}}}}
\end{align*}

We now need to study the Lipschitz smoothness of our function (modifying
the random matrix). We recall the original random-variable we wish
to bound $\inf_{v\in\mathbb{R}^{p}}\|R\bar{W}\bar{\Sigma}^{1/2}v\|_{2}$
where we treat this as a function of a matrix $M$ as
\[
h(M)=\sup_{v\in V(r)}\left(1-\frac{\|RM\bar{\Sigma}^{1/2}v\|_{2}}{\|\pi\|_{2}n^{1/2}}\right)\;.
\]

\begin{align*}
h(M)-h(M') & =\frac{1}{\|\pi\|_{2}n^{1/2}}\left(\sup_{v\in V(r)}\{-\|RM\bar{\Sigma}^{1/2}v\|_{2}\}-\sup_{v\in V(r)}\{-\|RM'\bar{\Sigma}^{1/2}v\|_{2}\}\right)\\
 & \le\frac{1}{\|\pi\|_{2}n^{1/2}}\sup_{v\in\mathcal{V}(r)}\|R(M-M')\bar{\Sigma}^{1/2}v\|_{2}\\
 & \le\frac{\|R\|_{2}}{\|\pi\|_{2}n^{1/2}}\left(\sup_{v\in\mathcal{V}(r)}\|\bar{\Sigma}^{1/2}v\|\right)\|M-M'\|_{2}\\
 & \le n^{-1/2}\|M-M'\|_{2}\;,
\end{align*}
where on the second line we use that there exists $\hat{v}=\arg\max_{v\in V(r)}\{-\|RM\bar{\Sigma}^{1/2}v\|_{2}$,
on the third line we use the sub-multiplactivity of $\|\cdot\|_{2}$,
and on the fourth $\|R\|_{2}=\|\pi\|_{2}$. Overall we have
\[
\frac{h(M)-h(M')}{\|M-M'\|_{F}}\le\frac{1}{n^{1/2}}\;,
\]
showing Lipschitz continuity with respect to the $\ell_{2}$ norm.
Replacing $M$ with the isotropic Gaussian random matrices $W$ and
$W'$ we have
\begin{align}
P\left[\big|M(r,W)-\mathbb{E}[M(r,W)]\big|\ge\frac{t(r)}{2}\right] & \le2\exp(-2^{-3}nt^{2}(r))\\
\implies P\left[M(r,W)\ge t'(r)\right] & \le2\exp(-2^{-1}3^{-2}nt'{}^{2}(r))\label{eq:mrw}
\end{align}
where in the second line we use $t'(r)=3t(r)/2$.

As in \cite{Yu2010}, we now apply a peeling argument, noting that
$t'(r)$ is bounded from below by $\mu=3\omega/2$. Let $\mathcal{V}':=\{v\in\mathbb{R}^{p}\:|\:\|\bar{\Sigma}^{1/2}v\|_{2}=1\}$
play a role similar to $\mathcal{V}(r)$, but without the $\ell_{1}$
norm constraining $\|v\|_{1}\le r$. We now assess the probability
of the random variable $n^{-1/2}\|R\bar{W}v\|_{2}$ being in a shell
of a given radius. Define the shells
\[
S_{m}:=\{v\in A\:|\:2^{m-1}\mu\le t'(\|v\|_{1})\le2^{m}\mu\}\;,\;m\in\mathbb{Z}_{+}\;,
\]
where we use $\|v\|_{1}=r$ to define the size of each shell, mapped
via $t'(r)$. Let us define $M(v,W):=1-n^{-1/2}\|R\bar{W}v\|_{2}$.
We now consider a given shell $m$, and examine vectors $v\in S_{m}$,
to check whether $M(v,W)\ge2t'(\|v\|_{1})$. If this occurs, then
we know $M(v,W)\ge2(2^{m-1})\mu=2^{m}\mu$. On the other hand for
any $v\in S_{m}$ we know $t'(\|v\|_{1})\le2^{m}\mu$ from the definition
of the shell. Consider the event:
\[
\mathcal{A}_{m}:=\{\exists v\in S_{m}\:|\:M(v,W)>2t'(\|v\|_{1})\}
\]
and the event 
\[
\mathcal{A}:=\{\exists v\in A\:|\:M(v,W)\ge2t'(\|v\|_{1})\}\;,
\]
we can bound $P[\mathcal{A}]\le\sum_{m=1}^{\infty}P[\mathcal{A}_{m}]$.
Writing $\mathcal{A}_{m}$ in terms of the maxima of $M(v,X)$ over
the $v\in S_{m}$ we have
\begin{align*}
P[\mathcal{A}_{m}] & =P[\sup_{t'(\|v\|_{1})\le2^{m}\mu}M(v,W)\ge2^{m}\mu]\\
 & =P[\sup_{\|v\|_{1}\le t'{}^{-1}(2^{m}\mu)}M(v,W)\ge2^{m}\mu]
\end{align*}
We already have a bound on the above given in (\ref{eq:mrw}), where
we use $r=t'{}^{-1}(2^{m}\mu)$, thus we find
\begin{align*}
P[\text{\ensuremath{\mathcal{A}}}] & \le P[\cup_{m=1}^{\infty}\mathcal{A}_{m}]\\
 & \le2\sum_{m=1}^{\infty}\exp(-2^{-1}3^{-2}n\{t'(t'{}^{-1}(2^{m}\mu))\}^{2})\\
 & =2\sum_{m=1}^{\infty}\exp(-2^{-1}3^{-2}n2^{2m}\mu^{2})\\
 & =\frac{2\exp(-3^{-2}2n\mu^{2})}{1-\exp(-3^{-2}2n\mu^{2})}\\
 & =\frac{2\exp(-2^{-1}\omega n)}{1-\exp(-2^{-1}\omega n)}
\end{align*}
We can thus conclude that $P[\mathcal{A}^{c}]\ge1-3\exp(-3^{2}2^{-1}\omega n)$.
And for $\|\bar{\Sigma}^{1/2}v\|_{2}=1$ we have 
\begin{align*}
1-\frac{\|R\bar{W}v\|_{2}}{\sqrt{n}} & \le2t'(\||v\|_{1})\\
 & =\omega+9\|\pi\|_{2}\|\Sigma\|_{\infty}^{1/2}\delta(n,p)\||v\|_{1}\\
\implies\frac{\|R\bar{W}v\|_{2}}{\sqrt{n}} & \ge(1-\omega)-9\|\pi\|_{2}\|\Sigma\|_{\infty}^{1/2}\delta(n,p)\||v\|_{1}\;.
\end{align*}
And thus, for any $v\in\mathbb{R}^{p}$ we have
\[
\frac{1}{\sqrt{n}}\frac{\|\tilde{X}v\|_{2}}{\|\pi\|_{2}}\ge(1-\omega)\|\Sigma^{1/2}v\|_{2}-9\|\Sigma\|_{\infty}^{1/2}\delta(n,p)\|v\|_{1}\;.
\]
as required.

\end{proof}

\subsection{Proof of Proposition \ref{prop:oracle}}

\begin{lemma}{Lambda choice under rotation \label{lemma:lambda-oracle}}

Let $\tilde{X}=RX$ and $\tilde{e}=Re$. Now, choose $\tilde{\lambda}_{n}\ge2c^{-1/2}K_{X}K_{u}\|\pi\|_{2}\delta(n,p^{\tau})$
then for all $n\ge c^{-1}(q+1)^{2}\log(p^{\tau})$ we have $\mathcal{E}_{\tilde{\lambda}}=\{2n^{-1}\|\tilde{X}^{\top}\tilde{e}\|_{\infty}\le\tilde{\lambda}_{n}\}$
holds in probability at least $1-2p^{1-\tau}$. 

\end{lemma}

\begin{proof}

Note that $Re=RLu=u$, and thus we need to construct a deviation bound
for $2n^{-1}\|X^{\top}R^{\top}u\|_{\infty}$ for which we can apply
Lemma \ref{lemma:tail} to obtain
\begin{align*}
P\left[\frac{2}{n}\|X^{\top}R^{\top}u\|_{\infty}\ge\lambda\right] & \le2p\exp\left(-c\min\left(\frac{\lambda^{2}n^{2}}{4K_{X}^{2}K_{u}^{2}\lVert R\rVert_{F}^{2}},\frac{\lambda n}{2K_{X}K_{u}\lVert R\rVert_{2}}\right)\right)\;.
\end{align*}
Further, using Lemma \ref{lemma:ma_comparison}, if 
\begin{equation}
\lambda\le\frac{2K_{X}K_{u}\|\pi\|_{2}}{q+1}\label{eq:lambda_oracle_choice}
\end{equation}
then we can use the Gaussian decay, to find
\[
P\left[\frac{2}{n}\|X^{\top}R^{\top}u\|_{\infty}\ge\lambda\right]\le2p\exp\left(\frac{-c\lambda^{2}n}{4K_{X}^{2}K_{u}^{2}\|\pi\|_{2}^{2}}\right)\;.
\]
Suggesting a choice of 
\[
\tilde{\lambda}\ge\frac{2K_{X}K_{u}\|\pi\|_{2}}{c^{1/2}}\delta(n,p^{\tau})\;.
\]
which satisfies Eq. \ref{eq:lambda_oracle_choice} for all 
\[
n\ge\frac{(q+1)^{2}}{c}\log(p^{\tau})\;,
\]
and leads to 
\[
P\left[\frac{2}{n}\|X^{\top}R^{\top}u\|_{\infty}\ge\lambda\right]\le1-2p^{1-\tau}\;.
\]
\end{proof}

\begin{proof}{Proposition \ref{prop:oracle} - Oracle Inequality}

We can construct a basic inequality post rotation akin to Eq. \ref{eq: rev3},
such that, where $\tilde{e}=Re=RLu=u$.
\[
0\le\frac{1}{n}\|\tilde{X}\hat{\Delta}\|_{2}^{2}\le\frac{2}{n}\lVert\tilde{X}^{\top}\tilde{e}\rVert_{\infty}\lVert\hat{\Delta}\rVert_{1}+2\tilde{\lambda}_{n}\{\lVert\beta_{\mathcal{S}}\rVert_{1}-\lVert\beta_{\mathcal{S}}+\hat{\Delta}_{\mathcal{S}}\rVert_{1}-\lVert\hat{\Delta}_{\mathcal{S}^{\perp}}\rVert_{1}+\|\beta_{\mathcal{S}^{\perp}}\|_{1}\}\;,
\]
and note that we have here accounted for
$\|\beta_{\mathcal{S}^{\perp}}\|_{1}$ possibly not zero. We then
have
\[
\frac{1}{n}\lVert\tilde{X}\hat{\Delta}\rVert_{2}^{2}\le\frac{2}{n}\lVert X^{\top}R^{\top}u\rVert_{\infty}\lVert\hat{\Delta}\rVert_{1}+2\tilde{\lambda}_{n}\{\lVert\beta_{\mathcal{S}}\rVert_{1}-\lVert\beta_{\mathcal{S}}+\hat{\Delta}_{\mathcal{S}}\rVert_{1}-\lVert\hat{\Delta}_{\mathcal{S}^{\perp}}\rVert_{1}+\|\beta_{\mathcal{S}^{\perp}}\|_{1}\}\;.
\]
conditioning on $\mathcal{E}_{\tilde{\lambda}}$ with choice of $\tilde{\lambda}_{n}$
from Lemma \ref{lemma:lambda-oracle}, we get
\[
\frac{1}{n}\lVert X\hat{\Delta}\rVert_{2}^{2}\le\tilde{\lambda}_{n}\big(3\|\hat{\Delta}_{\mathcal{S}}\|_{1}-\lVert\hat{\Delta}_{\mathcal{S}^{\perp}}\rVert_{1}+2\lVert\beta_{\mathcal{S}^{\perp}}\rVert_{1}\big)\;.
\]
Noting left is bounded away from zero gives

\begin{align*}
3\|\hat{\Delta}_{\mathcal{S}}\|_{1}-\lVert\hat{\Delta}_{\mathcal{S}^{\perp}}\rVert_{1}+2\lVert\beta_{\mathcal{S}}\rVert_{1} & \ge0\\
3\|\hat{\Delta}_{\mathcal{S}}\|_{1}+2\lVert\beta_{\mathcal{S}}\rVert_{1} & \ge\lVert\hat{\Delta}_{\mathcal{S}^{\perp}}\rVert_{1}
\end{align*}
Looking at the square of $\ell_{1}$ and upper bounding in light of
above gives: 
\begin{align*}
\|\hat{\Delta}\|_{1}^{2} & =\|\hat{\Delta}_{\mathcal{S}}+\hat{\Delta}_{\mathcal{S}^{\perp}}\|_{1}^{2}\\
 & =(\|\hat{\Delta}_{\mathcal{S}}\|_{1}+\|\hat{\Delta}_{\mathcal{S}^{\perp}}\|_{1})^{2}\\
 & \le(\|\hat{\Delta}_{\mathcal{S}}\|_{1}+3\|\hat{\Delta}_{\mathcal{S}}\|_{1}+2\lVert\beta_{\mathcal{S}^{\perp}}\rVert_{1})^{2}\\
 & =(4\|\hat{\Delta}_{\mathcal{S}}\|_{1}+2\lVert\beta_{\mathcal{S}^{\perp}}\rVert_{1})^{2}\\
 & \le2^{5}|\mathcal{S}|\|\hat{\Delta}\|_{2}^{2}+2^{3}\lVert\beta_{\mathcal{S}^{\perp}}\rVert_{1}^{2}
\end{align*}
From Lemma \ref{lemma:raskutti}, we have for all $\hat{\Delta}\in\mathbb{R}$
\[
\frac{1}{\sqrt{n}}\frac{\|\tilde{X}\hat{\Delta}\|_{2}}{\|\pi\|_{2}}\ge(1-\omega)\kappa\|\hat{\Delta}\|_{2}-9\|\Sigma\|_{\infty}^{1/2}\delta(n,p)\|\hat{\Delta}\|_{1}
\]
as long as $\mathcal{E}_{\mathrm{RE}}(\Sigma^{1/2};\kappa,\infty)$
on $\mathbb{R}^{p}$. Thus
\[
\frac{1}{n}\lVert\tilde{X}\hat{\Delta}\rVert_{2}^{2}\ge[\underset{c_{a}}{\underbrace{(1-\omega)\kappa\|\pi\|_{2}}}\|\hat{\Delta}\|_{2}-\underset{c_{b}}{\underbrace{9\|\Sigma\|_{\infty}^{1/2}\|\pi\|_{2}}}\delta(n,p)\|\hat{\Delta}\|_{1}]^{2}\;,
\]
where we treat $c_{a},c_{b}$ as constants with respect to $\|\hat{\Delta}\|_{2}.$We
thus have
\begin{align*}
\frac{1}{n}\lVert\tilde{X}\hat{\Delta}\rVert_{2}^{2} & \ge c_{a}^{2}\|\hat{\Delta}\|_{2}^{2}-2c_{a}c_{b}\|\hat{\Delta}\|_{2}\delta(n,p)\|\hat{\Delta}\|_{1}+c_{b}^{2}\delta^{2}(n,p)\|\hat{\Delta}\|_{1}^{2}\\
 & \ge c_{a}^{2}\|\hat{\Delta}\|_{2}^{2}-2c_{a}c_{b}\|\hat{\Delta}\|_{2}\delta(n,p)\|\hat{\Delta}\|_{1}\\
 & \ge c_{a}^{2}\|\hat{\Delta}\|_{2}^{2}-2c_{a}c_{b}\|\hat{\Delta}\|_{2}\delta(n,p)\sqrt{2^{5}|\mathcal{S}|\|\hat{\Delta}\|_{2}^{2}+2^{3}\lVert\beta_{\mathcal{S}^{\perp}}\rVert_{1}^{2}}\\
 & \ge c_{a}^{2}\|\hat{\Delta}\|_{2}^{2}-2c_{a}c_{b}\|\hat{\Delta}\|_{2}\delta(n,p)\{2^{3}\sqrt{|\mathcal{S}|}\|\hat{\Delta}\|_{2}+2^{2}\lVert\beta_{\mathcal{S}^{\perp}}\rVert_{1}\}\\
 & \ge\|\hat{\Delta}\|_{2}^{2}c_{a}\underset{c_{d}}{\underbrace{\left(c_{a}-2^{4}c_{b}\sqrt{|\mathcal{S}|}\delta(n,p)\right)}}-\underset{c_{e}}{\underbrace{2^{3}c_{a}c_{b}\delta(n,p)}}\lVert\beta_{\mathcal{S}^{\perp}}\rVert_{1}\|\hat{\Delta}\|_{2}
\end{align*}
note on line two we ignore the last term as it is strictly positive.
On line three we apply the bound we have on $\|\hat{\Delta}\|_{1}^{2}$.
On line 4, we use the triangle inequality on the square root term.
On line 5 we gather in terms of $\|\hat{\Delta}\|_{2}.$ Now we combine
with the upper bound
\begin{align*}
c_{a}c_{d}\|\hat{\Delta}\|_{2}^{2} & \le\tilde{\lambda}_{n}\big(3\|\hat{\Delta}_{\mathcal{S}}\|_{1}-\lVert\hat{\Delta}_{\mathcal{S}^{\perp}}\rVert_{1}+2\lVert\beta_{\mathcal{S}^{\perp}}\rVert_{1}\big)+c_{e}\lVert\beta_{\mathcal{S}^{\perp}}\rVert_{1}\|\hat{\Delta}\|_{2}\\
 & \le3\tilde{\lambda}_{n}\sqrt{|\mathcal{S}|}\|\hat{\Delta}\|_{2}+2\tilde{\lambda}_{n}\lVert\beta_{\mathcal{S}^{\perp}}\rVert_{1}+c_{e}\lVert\beta_{\mathcal{S}^{\perp}}\rVert_{1}\|\hat{\Delta}\|_{2}\\
 & \le\underset{c_{f}}{\underbrace{\big(3\tilde{\lambda}_{n}\sqrt{|\mathcal{S}|}+c_{e}\lVert\beta_{\mathcal{S}^{\perp}}\rVert_{1}\big)}}\|\hat{\Delta}\|_{2}+\underset{c_{g}}{\underbrace{2\tilde{\lambda}_{n}\lVert\beta_{\mathcal{S}^{\perp}}\rVert_{1}}}\\
0 & \ge c_{a}c_{d}\|\hat{\Delta}\|_{2}^{2}-c_{f}\|\hat{\Delta}\|_{2}-c_{g}
\end{align*}
Consider the roots $(2c_{a}c_{d})^{-1}(c_{f}\pm\sqrt{c_{f}^{2}+4c_{a}c_{d}c_{g}})$.
Assuming $c_{d}>0$ implies only one positive root. For simplicity
assume $c_{d}>c_{a}/2>0$ which occurs if $2^{3}c_{b}\sqrt{|\mathcal{S}|}\delta(n,p)<c_{a}^ {}$,
which is implied by the sample size condition
\begin{align*}
n & \ge\frac{3^{4}2^{6}\|\Sigma\|_{\infty}|\mathcal{S}|\log p}{(1-\omega)^{2}\kappa^{2}}\;.
\end{align*}
Thus, under sufficient samples, we obtain\footnote{Look at the square of the positive root in Eq. \ref{eq:roots} and
multiply by 2 to avoid the cross term.} 

\begin{align*}
\|\hat{\Delta}\|_{2}^{2} & \le\frac{2}{4c_{a}^{2}c_{d}^{2}}\big(c_{f}^{2}+c_{f}^{2}+4c_{d}c_{a}c_{g}\big)\\
 & =\frac{(3\tilde{\lambda}_{n}\sqrt{|\mathcal{S}|}+c_{e}\lVert\beta_{\mathcal{S}^{\perp}}\rVert_{1})^{2}}{c_{a}^{2}c_{d}^{2}}+\frac{2\overset{c_{g}}{\overbrace{\tilde{\lambda}_{n}\lVert\beta_{\mathcal{S}^{\perp}}\rVert_{1}}}}{c_{a}c_{d}}\\
= & \frac{2.3^{2}\tilde{\lambda}_{n}^{2}|\mathcal{S}|+2(2^{3}c_{a}c_{b}\delta(n,p))^{2}\lVert\beta_{\mathcal{S}^{\perp}}\rVert_{1}^{2}}{c_{a}^{2}c_{d}^{2}}+\frac{2\tilde{\lambda}_{n}\lVert\beta_{\mathcal{S}^{\perp}}\rVert_{1}}{c_{a}c_{d}}\\
= & \frac{18\tilde{\lambda}_{n}^{2}|\mathcal{S}|}{c_{a}^{2}c_{d}^{2}}+\frac{2^{4}3^{4}\|\Sigma\|_{\infty}^{2}\|\pi\|_{2}^{2}\delta^{2}(n,p)\lVert\beta_{\mathcal{S}^{\perp}}\rVert_{1}^{2}}{c_{d}^{2}}+\frac{2\tilde{\lambda}_{n}\lVert\beta_{\mathcal{S}^{\perp}}\rVert_{1}}{c_{a}c_{d}}\\
\le & \frac{2^{2}.18\tilde{\lambda}_{n}^{2}|\mathcal{S}|}{(1-\omega)^{4}\kappa^{4}\lVert\pi\rVert_{2}^{4}}+\frac{2^{2}2^{4}3^{4}\|\Sigma\|_{\infty}^{2}\|\pi\|_{2}^{2}\delta^{2}(n,p)\lVert\beta_{\mathcal{S}^{\perp}}\rVert_{1}^{2}}{(1-\omega)^{2}\kappa^{2}\lVert\pi\rVert_{2}^{2}}+\frac{2^{2}\tilde{\lambda}_{n}\lVert\beta_{\mathcal{S}^{\perp}}\rVert_{1}}{(1-\omega)^{2}\kappa^{2}\lVert\pi\rVert_{2}^{2}}\\
= & \frac{2^{3}.3^{2}\tilde{\lambda}_{n}^{2}|\mathcal{S}|}{(1-\omega)^{4}\kappa^{4}\lVert\pi\rVert_{2}^{4}}+\frac{2^{2}\tilde{\lambda}_{n}\lVert\beta_{\mathcal{S}^{\perp}}\rVert_{1}}{(1-\omega)^{2}\kappa^{2}\lVert\pi\rVert_{2}^{2}}+\frac{2^{6}3^{4}\|\Sigma\|_{\infty}^{2}\delta^{2}(n,p)\lVert\beta_{\mathcal{S}^{\perp}}\rVert_{1}^{2}}{(1-\omega)^{2}\kappa^{2}}
\end{align*}
\end{proof}

\subsection{Proofs for Proposition \ref{prop:phi_err}}

\begin{proof}

We will use the notation $e'_{n}$ to indicate a vector of length
$m=n-q$ containing elements $e'_{n}=(e_{q+1},\ldots,e_{n})^{\top}$.
We can write the estimated residuals as $\hat{e}'_{n}=e'_{n}+z'_{n}$
where $z'_{n}=(z_{q+1},\ldots,z_{n})$ represents the impact of the
mis-specified expectation, i.e. $z'_{n}=(\Delta^{\top}x_{q+1},\ldots,\Delta^{\top}x_{n})^{\top}$.
For convenience, let us consider the $q\times q$ dimensional matrices
\[
\hat{S}=m^{-1}\hat{E}^{\top}\hat{E}\quad;\quad Q_{m}=\hat{S}^{-1}\;,
\]
 where $\hat{E}=E+Z$, and
\[
E=\left(\begin{array}{ccc}
e_{q} & \ldots & e_{1}\\
\vdots &  & \vdots\\
e_{n-1} & \ldots & e_{n-q}
\end{array}\right)\quad;\quad Z=\left(\begin{array}{ccc}
z_{q} & \ldots & z_{1}\\
\vdots &  & \vdots\\
z_{n-1} & \ldots & z_{n-q}
\end{array}\right)\;,
\]
represent the lagged variables on which we project $\hat{e}'_{n}$.
Recall $\hat{\phi}=Q_{m}\hat{E}^{\top}\hat{e}'_{n}$, which we can
further expand as
\begin{align*}
\hat{\phi} & =m^{-1}Q_{m}\hat{E}^{\top}(e'_{n}+z'_{n})\\
 & =Q_{m}\big(m^{-1}\hat{E}^{\top}e'_{n}\big)+Q_{m}\big(m^{-1}\hat{E}^{\top}z'_{n}\big)
\end{align*}
Now writing $e'_{n}=E\phi+u'_{n}$ where $\phi=(\phi_{1},\ldots\phi_{p})^{\top}$
we obtain
\begin{align*}
\hat{\phi} & =Q_{m}(m^{-1}\hat{E}^{\top}E)\phi+Q_{m}\underset{=:v_{1}}{\underbrace{(m^{-1}\hat{E}^{\top}u'_{n})}}+\underset{=:v_{2}}{Q_{m}\underbrace{(m^{-1}\hat{E}^{\top}z'_{n})}}\\
 & =Q_{m}(m^{-1}\hat{E}^{\top}E)\phi+Q_{m}(v_{1}+v_{2})\;,
\end{align*}
writing in terms of $\hat{E}$ gives
\begin{align*}
\hat{\phi} & =Q_{m}m^{-1}\big[\hat{E}^{\top}(\hat{E}-Z)\big]\phi+Q_{m}(v_{1}+v_{2})\\
\implies\hat{\phi}-\phi & =-Q_{m}\underset{=:V}{\underbrace{(m^{-1}\hat{E}^{\top}Z)}}\phi+Q_{m}(v_{1}+v_{2})\;.
\end{align*}
We consider the maximum error incurred, and the decomposition

\begin{align*}
\|\hat{\phi}-\phi\|_{\infty} & \le\|Q_{m}V\phi\|_{\infty}+\|Q_{m}(v_{1}+v_{2})\|_{\infty}\\
 & \le|Q_{m}|_{\infty}|V|_{\infty}\|\phi\|_{\infty}+|Q_{m}|_{\infty}(\|v_{1}\|_{\infty}+\|v_{2}\|_{\infty})\;,
\end{align*}
where $|Q_{m}|_{\infty}:=\max_{j}\|Q_{m;j\cdot}\|_{1}$.

\subsubsection*{\emph{Bounding $Q_{m}$}}

Note that the $ij$th entry of $\hat{S}$ is given by
\begin{align*}
\hat{S}_{ij} & =\frac{1}{m}\sum_{t=q}^{m+q}\hat{e}_{t-(q-i)}\hat{e}_{t-(q-j)}\\
 & =\frac{1}{m}\sum_{t'=1}^{m}\hat{e}_{t'-i}\hat{e}_{t'-j}\;,
\end{align*}
where $t'=t-q$ for $i,j\in[q]$. Let $S_{0}=\mathbb{E}[\hat{S}]$,
expanding in terms of the above we find
\begin{align*}
S_{0;ij} & =\mathbb{E}\left[\frac{1}{m}\sum_{t'=1}^{m}e_{t'-i}e_{t'-j}\right]+\mathbb{E}\left[\frac{1}{m}\sum_{t'=1}^{m}z_{t'-i}z_{t'-j}\right]\\
 & =\Gamma_{ij}+\boldsymbol{1}_{i=j}\mathbb{E}\left[\frac{1}{m}\sum_{t'=1}^{m}z_{t'}^{2}\right]\;,
\end{align*}
where $\boldsymbol{1}_{i=j}=0$ if $i\ne j$ and one otherwise. Further
note
\begin{align*}
z_{t}^{2} & =(\Delta^{\top}\Sigma^{1/2}w_{t})^{2}\\
 & =\Delta^{\top}\Sigma^{1/2}w_{t}(\Delta^{\top}\Sigma^{1/2}w_{t})^{\top}\\
 & =\Delta^{\top}\Sigma^{1/2}w_{t}w_{t}^{\top}(\Sigma^{1/2})^{\top}\Delta
\end{align*}
and thus $\mathbb{E}\left[m^{-1}\sum_{t'=1}^{m}z_{t'}^{2}\right]=\Delta^{\top}\Sigma\Delta=\|\Sigma^{1/2}\Delta\|_{2}^{2}$.

Now, let $D:=\hat{S}-S_{0}$ and assume that $\|D\|\le\epsilon$,
we desire to find a bound on $Q_{m}$. Expanding, we have
\begin{align*}
Q_{m} & =(S_{0}+D)^{-1}\\
 & =(S_{0}S_{0}^{-1}S_{0}+DS_{0}^{-1}S_{0})^{-1}\\
 & =\{(I+DS_{0}^{-1})S_{0}\}^{-1}\\
 & =S_{0}^{-1}(I+DS_{0}^{-1})^{-1}\;.
\end{align*}
If $|DS_{0}^{-1}|_{\infty}<1$ then we have the expansion
\begin{align*}
|Q_{m}|_{\infty} & =|S_{0}^{-1}\sum_{k=0}^{\infty}(-1)^{k}(DS_{0}^{-1})^{k}|_{\infty}\\
 & \le|S_{0}^{-1}|_{\infty}|\sum_{k=0}^{\infty}(-1)^{k}(DS_{0}^{-1})^{k}|_{\infty}\\
 & \le\frac{|S_{0}^{-1}|_{\infty}}{1-|DS_{0}^{-1}|_{\infty}}\;.
\end{align*}
Note that $|DS_{0}^{-1}|_{\infty}\le|D|_{\infty}|S_{0}^{-1}|_{\infty}\le q\|D\|_{\infty}|S_{0}^{-1}|_{\infty}$.
Thus if we condition on the event
\begin{equation}
\mathcal{E}_{D}:=\left\{ \|D\|_{\infty}<\frac{1}{2q|S_{0}^{-1}|_{\infty}}\right\} \;,\label{eq:EW}
\end{equation}
we have $|DS_{0}^{-1}|_{\infty}\le1/2$ and $|Q_{m}|_{\infty}\le2|S_{0}^{-1}|_{\infty}$.
We now work to show $P[\mathcal{E}_{D}]\rightarrow1$ by decomposing
$\|D\|_{\infty}$ into three parts, we will consider that if any of
these exceed some set threshold then the sum will also exceed this
threshold. Let us define
\begin{align*}
D_{E} & :=\frac{1}{m}E^{\top}E-\Gamma_{q}\\
D_{EZ} & :=\frac{1}{m}E^{\top}Z\\
D_{Z} & :=\frac{1}{m}Z^{\top}Z-\|\Sigma^{1/2}\Delta\|_{2}^{2}I_{q}\;,
\end{align*}
where we note $\Gamma_{q}\in\mathbb{R}^{q\times q}$ is simply a subset
of $\Gamma\in\mathbb{R}^{n\times n}$, and note $\|D\|_{\infty}\le\|D_{E}\|_{\infty}+2\|D_{EZ}\|_{\infty}+\|D_{Z}\|_{\infty}$.
Define events $\mathcal{E}_{D_{E}}:=\{\|D_{E}\|_{\infty}<1/2q|S_{0}^{-1}|_{\infty}\}$
for each of the three terms, i.e. we have $\mathcal{E}_{D_{E}},\mathcal{E}_{D_{EZ}},\mathcal{E}_{D_{Z}}$
defined analogously to (\ref{eq:EW}). Note that $\mathcal{E}_{D}^{c}\subset\{\mathcal{E}_{D_{E}}^{c}\cup\mathcal{E}_{D_{EZ}}^{c}\cup\mathcal{E}_{D_{Z}}^{c}\}$
thus $P[\mathcal{E}_{D}^{c}]\le P[\mathcal{E}_{D_{E}}^{c}\cup\mathcal{E}_{D_{EZ}}^{c}\cup\mathcal{E}_{D_{Z}}^{c}]$.
Writing $E=L'U$, where $U$ is an array like $E$, but with white-noise
terms $\{u_{t}\}$ and $L'$ is the $m\times m$ sub-matrix of $L$
for obtaining the relevant time-indices contained in $E$. We can
now apply Lemma \ref{lemma:hanson_wright}, Lemma \ref{lemma:tail},
and Lemma \ref{lemma:hanson_wright} respectively, to obtain:

\begin{align*}
P[\|D_{E}\|_{\infty}>\epsilon] & \le2q^{2}\exp\left(-c\min\left(\frac{\epsilon^{2}m^{2}}{K_{u}^{4}\lVert L'{}^{\top}L'\rVert_{F}^{2}},\frac{\epsilon m}{K_{u}^{2}\lVert L'{}^{\top}L'\rVert_{2}}\right)\right)\\
 & \le2q^{2}\exp\left(-c\min\left(\frac{\epsilon^{2}m}{K_{u}^{4}\|\Gamma'\|_{2}^{2}},\frac{\epsilon m}{K_{u}^{2}\|\Gamma'\|_{2}}\right)\right)
\end{align*}
where we let $\Gamma'=\Gamma/\sigma_{u}^{2}$, and note $\|L'{}^{\top}L'\|_{F}^{2}=\|\Gamma'_{m}\|_{F}^{2}\le m\|\Gamma'\|_{2}^{2}$
.
\begin{align}
P[\|D_{EZ}\|_{\infty}>\epsilon/2] & \le2q^{2}\exp\left(-c\min\left(\frac{\epsilon^{2}m^{2}}{4K_{u}^{2}K_{z}^{2}\lVert L'\rVert_{F}^{2}},\frac{\epsilon m}{2K_{u}K_{z}\lVert L'\rVert_{2}}\right)\right)\nonumber \\
 & \le2q^{2}\exp\left(-c\min\left(\frac{\epsilon^{2}m}{4K_{u}^{2}K_{z}^{2}v^{2}},\frac{\epsilon m}{2K_{u}K_{z}v}\right)\right)\nonumber \\
 & \le2q^{2}\exp\left(-c\min\left(\frac{\epsilon^{2}m}{K_{u}^{2}K_{x}^{2}\|\Delta\|_{2}^{2}v^{2}},\frac{\epsilon m}{K_{u}K_{x}\|\Delta\|_{2}v}\right)\right)\label{eq:Wez}
\end{align}
where we use $\|L'\|_{F}^{2}=mv^{2}$ and on the last line we use
$K_{z}\le\|\Delta\|_{2}K_{x}$ and absorb the factor of $1/4$ into
the constant $c$.

\begin{align}
P[\|D_{Z}\|_{\infty}>\epsilon] & \le2q^{2}\exp\left(-c\min\left(\frac{\epsilon^{2}m}{K_{z}^{4}},\frac{\epsilon m}{K_{z}^{2}}\right)\right)\nonumber \\
 & \le2q^{2}\exp\left(-c\min\left(\frac{\epsilon^{2}m}{K_{x}^{4}\|\Delta\|_{2}^{4}},\frac{\epsilon m}{K_{x}^{2}\|\Delta\|_{2}^{2}}\right)\right)\;.\label{eq:eqWZZ}
\end{align}
Consider $\|\Gamma'\|_{2}\ge v\ge1$, so that Comparing the tails,
we have 
\[
\min\left[\frac{1}{K_{u}^{4}\|\Gamma'\|_{2}^{2}},\frac{1}{K_{u}^{2}K_{x}^{2}\|\Delta\|_{2}^{2}v^{2}},\frac{1}{K_{x}^{4}\|\Delta\|_{2}^{4}}\right]=\frac{1}{K_{u}^{4}\|\Gamma'\|_{2}^{2}}
\]
for all $\|\Delta\|_{2}\le K_{u}/K_{x}$, where we note that $\|\Gamma'\|_{2}\ge v\ge1$
. We can now combine the bounds (applying the union bound) to find
\[
P\big[\{\|D_{E}\|_{\infty}>\epsilon\}\cup\{\|D_{EZ}\|_{\infty}>\epsilon\}\cup\{\|D_{Z}\|_{\infty}>\epsilon\}\big]\le6q^{2}\exp\left(-c\frac{\epsilon^{2}m}{K_{u}^{4}\|\Gamma'\|_{2}^{2}}\right)
\]
Now choose $\epsilon=c^{-1/2}K_{u}^{2}\|\Gamma'\|_{2}\delta(m,p^{\tau})$
and assuming $q\le p^{1/2}$ we obtain
\begin{align*}
P[\mathcal{E}_{D_{E}}^{c}\cup\mathcal{E}_{D_{EZ}}^{c}\cup\mathcal{E}_{D_{Z}}^{c}] & \le6p^{1-\tau}\\
\implies P[\mathcal{E}_{D}] & \ge1-6p^{1-\tau}\;,
\end{align*}
as long as $\epsilon\le\min[K_{u}^{2}\|\Gamma'\|_{2},1/2q|S_{0}^{-1}|_{\infty}]$,
which is implied\footnote{Note: $|X|_{\infty}^{2}\le q\|X\|_{F}^{2}$ for a generic $X\in\mathbb{R}^{q\times q}$.
We therefore have $|S_{0}^{-1}|_{\infty}\le q/\{\gamma_{\min}(\Gamma_{q})+\|\Sigma^{1/2}\Delta\|_{2}^{2}\}\le q\gamma_{\min}^{-1}(\Gamma_{q}).$} by a choice of 
\[
n\ge\max\left[1+q,\frac{K_{u}^{4}}{\sigma_{u}^{4}}q^{4}\frac{\gamma_{\max}^{2}(\Gamma)}{\gamma_{\min}^{2}(\Gamma)}\right]\frac{\log(p^{\tau})}{c}\;,
\]
where we have used $2\delta(n,p^{\tau})\ge\delta(m,p^{\tau})$, which
for all $n>4q/3$, to simplify the condition from a bound on $m$
to $n$.

\subsubsection*{\emph{Bounding $V$}}

Let $V_{0}:=\mathbb{E}[V]=\mathbb{E}[m^{-1}(E^{\top}Z+Z^{\top}Z)]$
and thus $V_{0;ij}=\|\Sigma^{1/2}\Delta\|_{2}^{2}\boldsymbol{1}_{i=j}$,
we further have $|V|_{\infty}\le q\|V\|_{\infty}$, letting $D_{V}=V-V_{0}$
we have
\begin{align*}
P[\|D_{V}\|_{\infty}>\epsilon] & \le P[\|D_{EZ}\|_{\infty}>\epsilon]+P[\|D_{Z}\|_{\infty}>\epsilon]\\
 & \le4q^{2}\exp\left(-c\frac{\epsilon^{2}m}{K_{u}^{2}K_{x}^{2}\|\Delta\|_{2}^{2}v^{2}}\right)
\end{align*}
with a choice of $\epsilon=c_{}^{-1/2}vK_{u}K_{x}\|\Delta\|_{2}\delta(m,p^{\tau})$,
we then obtain
\[
P\big[\|V\|_{\infty}<\|\Sigma^{1/2}\Delta\|_{2}^{2}+c^{-1/2}vK_{u}K_{x}\|\Delta\|_{2}\delta(m,p^{\tau})\big]\ge1-4p^{1-\tau}\;,
\]
where we have assumed $K_{u}^{2}v^{2}\big/K_{x}^{2}\ge\|\Delta\|_{2}^{2}$.

\subsubsection*{\emph{Bounding $v_{1}$}}

Recall $v_{1}:=m^{-1}\hat{E}^{\top}u'$ and thus $v_{1;0}=\mathbb{E}[m^{-1}(E+Z)^{\top}u']=\mathbb{E}[m^{-1}E^{\top}u']$.
Now note that $u'$ is indexed in the future relative to $U$ (analogously
to $e'_{n})$, so when we consider
\[
[E^{\top}u']_{i}=[U^{\top}(L')^{\top}u']_{i}=\sum_{t=q+1}^{n-1}\sum_{l=1}^{m}U_{li}L_{tl}u_{t+1}\;,
\]
for every index $t$, we have that $\sum_{l=1}^{m}U_{li}L_{tl}$ references
only historic shock terms, i.e. doesn't involve $u_{t+1}$ in the
sum. We thus get $\mathbb{E}[E^{\top}u']=0$. Deviations $d_{v_{1}}:=v_{1}-v_{1;0}$
can be bound by applying Lemma \ref{lemma:tail}, similar to (\ref{eq:Wez})
but taking the union bound over $q$ rather than $q^{2}$ elements
to obtain
\[
P[\|d_{v_{1}}\|_{\infty}>\epsilon]\le2q\exp\left(-c\frac{\epsilon^{2}m}{K_{u}^{4}v^{2}}\right)
\]
choosing $\epsilon=c^{-1/2}K_{u}^{2}v\delta(m,p^{\tau})$ for $\tau>1$
gives
\[
P[\|v_{1}\|_{\infty}\le c^{-1/2}K_{u}^{2}v\delta(m,p^{\tau})]\ge1-2p^{1-\tau}\;.
\]

\subsubsection*{\emph{Bounding $v_{2}$}}

Recall $v_{2}:=m^{-1}\hat{E}^{\top}z'$ and thus $v_{2;0}=m^{-1}\mathbb{E}[E^{\top}z'+Z^{\top}z']=0$
as similar to the case with $v_{1;0}$ we have $z'$ is forward in
time relative to the variates in $E$ and $Z$. Again, applying Lemma
\ref{lemma:tail}, letting $d_{v_{2}}=v_{2}-v_{2;0}$, we find 
\[
P[\|d_{v_{2}}\|_{\infty}>\epsilon]\le2q\exp\left(-c_{}\frac{\epsilon^{2}m}{K_{u}^{2}K_{x}^{2}\|\Delta\|_{2}^{2}v^{2}}\right)
\]
we can choose 
\[
\epsilon=c_{}^{1/2}K_{u}K_{x}\|\Delta\|_{2}v\delta(m,p^{\tau})
\]
to obtain
\[
P\left[\|v_{2}\|_{\infty}\le c_{}^{1/2}K_{u}K_{x}\|\Delta\|_{2}v\delta(m,p^{\tau})\right]\ge1-2p^{1-\tau}\;.
\]

\subsubsection*{\emph{Combined Bound}}

Recall $|V|_{\infty}\le q\|V\|_{\infty}$, and thus 
\[
|V|_{\infty}\|\phi\|_{\infty}\le q\|\Sigma^{1/2}\Delta\|_{2}^{2}\|\phi\|_{\infty}+c^{-1/2}vK_{u}K_{x}\|\Delta\|_{2}q\delta(m,p^{\tau})\|\phi\|_{\infty}\;.
\]
Combining the results of $v_{1}$ and $v_{2}$, we have
\begin{align*}
\|v_{1}\|_{\infty}+\|v_{2}\|_{\infty} & \le c^{-1/2}K_{u}v\{K_{u}+K_{x}\|\Delta\|_{2}\}q\delta(m,p^{\tau})\;,
\end{align*}
Now, noting $|Q_{m}|_{\infty}\le2|S_{0}^{-1}|_{\infty}\le2q\gamma_{\min}^{-1}(\Gamma_{q})$,
we have 
\[
\|\hat{\phi}-\phi\|_{\infty}\le\frac{2q^{2}}{\gamma_{\min}(\Gamma_{q})}\left[\|\Sigma^{1/2}\Delta\|_{2}^{2}\|\phi\|_{\infty}+2c^{-1/2}K_{u}^{2}v\delta(m,p^{\tau})\right]
\]
in probability greater than $1-14p^{1-\tau}$ where we have assumed:
\begin{enumerate}
\item The error is bounded such that $\|\Delta\|_{2}\le K_{u}/K_{x}$
\item Sufficient samples:
\[
n\ge\max\left[1+q,\frac{K_{u}^{4}}{\sigma_{u}^{4}}q^{4}\frac{\gamma_{\max}^{2}(\Gamma)}{\gamma_{\min}^{2}(\Gamma)}\right]\frac{\log(p^{\tau})}{c}\;.
\]
\end{enumerate}
\end{proof}

\begin{proof}{Corollary \ref{cor:phi_bound}}

Using the initial LASSO error where we have assumed $\mathcal{E}_{RE}(\Sigma^{1/2},\kappa,3\alpha)$
over $\mathcal{S}=\mathcal{S}_{0}$, with $|\mathcal{S}_{0}|=s$ and
we further set $\alpha=3$ (sufficient for the LASSO), we obtain from
Eq. \ref{eq:lasso_l2_err} the first stage lasso error bound, conditional
on $\mathcal{E}_{\mathrm{RE}}(n^{-1/2}X;\kappa,\alpha)$, given by
\[
\|\hat{\Delta}\|_{2}\le3\frac{K_{X}K_{u}}{c^{1/2}\kappa}v\sqrt{\frac{s\log p^{\tau}}{n}}\;,
\]
which holds in probability greater than $1-2p^{-\tau}$, as long as
we have $n\ge2^{-2}c\log p^{\tau}$ and set $\lambda_{n}=\frac{K_{X}K_{u}}{c^{1/2}}v\delta(n,p^{\tau})$.
If we want to check the RE condition for the sub-Gaussian design,
we can apply Proposition \ref{prop:re_subgaussian}, which ensures
$\mathcal{E}_{\mathrm{RE}}(n^{-1/2}X;(1-\zeta_{1})\kappa,\alpha)$
holds, for convenience, let us choose $\zeta_{1}=1/2$. We thus have
\[
\|\hat{\Delta}\|_{2}\le6\frac{K_{X}K_{u}}{c^{1/2}\kappa}v\sqrt{\frac{s\log p^{\tau}}{n}}
\]
 in probability greater than $1-4p^{-\tau}$ under the previous choice
of $\lambda_{n}$, and as long as we have sufficient samples
\[
n\ge\frac{2^{4}K_{W}^{4}\alpha^{3}}{c}\max_{j\in\mathcal{S}}\|\Sigma^{1/2}b_{j}\|_{2}^{2}\frac{s}{\kappa^{2}}\log(p^{\tau})\;.
\]
We first check the condition $\|\hat{\Delta}\|_{2}\le K_{u}/K_{x}$
which is satisfied if
\[
n\ge\frac{3^{2}2^{2}K_{X}^{4}}{c\kappa^{2}}v^{2}s\log p^{\tau}\;.
\]
Noting that $\max_{j\in\mathcal{S}}\|\Sigma^{1/2}b_{j}\|_{2}^{2}\le\|\Sigma^{1/2}\|_{2}^{2}$
and $K_{X}\le K_{W}\|\Sigma^{1/2}\|_{2}$ we can combine the sample
size conditions to find that with
\[
n\ge\max\left[1+q,\frac{K_{u}^{4}}{\sigma_{u}^{4}}q^{4}\frac{\gamma_{\max}^{2}(\Gamma)}{\gamma_{\min}^{2}(\Gamma)},\frac{s}{\kappa^{2}}\|\Sigma^{1/2}\|_{2}^{4}K_{W}^{4}v^{2}\right]\frac{\log(p^{\tau})}{c}
\]
we have
\begin{align*}
\|\hat{\phi}-\phi\|_{\infty} & \le\frac{2q^{2}}{\gamma_{\min}(\Gamma_{q})}\left[\|\Sigma^{1/2}\Delta\|_{2}^{2}\|\phi\|_{\infty}+2c^{-1/2}K_{u}^{2}v\delta(m,p^{\tau})\right]\\
 & \le\frac{2q^{2}}{\gamma_{\min}(\Gamma_{q})}\bigg[\|\Sigma^{1/2}\|_{2}^{2}\frac{36K_{X}^{2}K_{u}^{2}}{c\kappa^{2}}v^{2}s\delta^{2}(n,p^{\tau})\|\phi\|_{\infty}+\frac{2K_{u}^{2}v}{c^{1/2}}\delta(m,p^{\tau})\bigg]\\
 & \le\frac{8K_{u}^{2}q^{2}v}{c^{1/2}\gamma_{\min}(\Gamma_{q})}\bigg[\|\Sigma^{1/2}\|_{2}^{2}\frac{9K_{X}^{2}}{c^{1/2}\kappa^{2}}vs\delta^{2}(n,p^{\tau})\|\phi\|_{\infty}+\delta(n,p^{\tau})\bigg]
\end{align*}
where on the last line we have used $2\delta(n,p^{\tau})\ge\delta(m,p^{\tau})$.
We thus have 
\begin{align*}
\|\hat{\phi}-\phi\|_{\infty} & \le\frac{8K_{u}^{2}q^{2}v}{c^{1/2}\gamma_{\min}(\Gamma_{q})}\delta(n,p^{\tau})\left[\|\Sigma^{1/2}\|_{2}^{4}\frac{3^{2}K_{W}^{2}}{c^{1/2}\kappa^{2}}vs\delta(n,p^{\tau})\|\phi\|_{\infty}+1\right]\\
 & \le\frac{2^{4}K_{u}^{2}q^{2}v}{c_{1}^{1/2}\gamma_{\min}(\Gamma_{q})}\delta(n,p^{\tau})\;,
\end{align*}
where the last line holds under the condition $\|\Sigma^{1/2}\|_{2}^{2}\frac{3^{2}K_{W}^{2}}{c^{1/2}\kappa^{2}}vs\delta(n,p^{\tau})\|\phi\|_{\infty}\le1$
implied by the sample size condition

\[
n\ge\frac{3^{4}K_{W}^{4}}{c\kappa^{4}}\|\Sigma^{1/2}\|_{2}^{4}\|\phi\|_{\infty}^{2}v^{2}s^{2}\log p^{\tau}\;.
\]
Combining the sample size conditions gives
\begin{align*}
n & \ge\max\left[1+q,\frac{K_{u}^{4}}{\sigma_{u}^{4}}q^{4}\frac{\gamma_{\max}^{2}(\Gamma)}{\gamma_{\min}^{2}(\Gamma)},\frac{K_{W}^{4}\|\Sigma^{1/2}\|_{2}^{4}}{\kappa^{2}}v^{2}s,\frac{K_{W}^{4}}{\kappa^{4}}\|\Sigma^{1/2}\|_{2}^{4}v^{2}s^{2}\|\phi\|_{\infty}^{2}\right]\frac{\log(p^{\tau})}{c'_{2}}\;.
\end{align*}
In the final presentation, we simplify the requirement to 
\[
n\ge c_{2}\kappa^{-2}q^{4}v^{2}s\max[1,\kappa^{-2}\|\Sigma^{1/2}\|_{2}^{4}s\|\phi\|_{\infty}^{2}]\log(p^{\tau})
\]
 where $c_{2}$ takes into account $K_{u},K_{W}$.

\end{proof}

\begin{proof}{Proof of Proposition \ref{prop:fgls}}

We consider the expansion in terms of rotation error $D=\hat{R}-R$
according to $\widehat{R}L=(R+D)L=I_{n}+DL$. The strategy is to relate
to previous arguments based on $R^{\top}R$ in relation to $\pi$.
We now consider the specific structure in $D^{\top}D$ akin to $R^{\top}R$
in Lemma \ref{lemma:ma_comparison}. Specifically, we have $\|D\|_{F}^{2}=\mathrm{tr}(D^{\top}D)\le n\|\hat{\pi}-\pi\|_{2}^{2}=n\|\Delta_{\phi}\|_{2}^{2}$
where $\Delta_{\phi}:=\hat{\phi}-\phi$, we further compute $\|D\|_{2}=\sqrt{\gamma_{\max}(D^{\top}D)}$.
Recall from earlier (Lemma \ref{lemma:ma_comparison}), we have $\|R^{\top}R\|_{2}\le\|\Sigma_{\mathrm{MA}(\pi)}\|_{2}$.
A similar argument can now be made on $D$, such that $\|D^{\top}D\|_{2}\le\|\Sigma_{\mathrm{MA}(\hat{\pi}-\pi)}\|_{2}$
where we will have $\Sigma_{\mathrm{MA}(\hat{\pi}-\pi),s,t}$ taking
the form of a Toeplitz matrix with the $|s-t|=k$th off-diagonal entry
given by
\begin{align*}
\gamma_{\mathrm{MA}(\hat{\pi}-\pi);k} & =\sum_{j=0}^{q-k}(\hat{\pi}-\pi)_{j}(\hat{\pi}-\pi)_{j+k}=\sum_{j=1}^{q-k}\Delta_{\phi,j}\Delta_{\phi,j+k}\;.
\end{align*}
We then bound the $\ell_{1}$ row-norm of $D^{\top}D$, and thus find
$\|D\|_{2}\le\sqrt{2q+1}\|\Delta_{\phi}\|_{2}.$ For choosing $\bar{\lambda}_{n}$
we recall we want the event $\mathcal{E}_{\bar{\lambda}}:=\{2n^{-1}\|(\hat{R}X)^{\top}\hat{R}Lu\|_{\infty}<\lambda_{n}\}$
to occur in high probability. We are thus interested in the matrix
\begin{align*}
\hat{R}^{\top}\hat{R}L & =(R+D)^{\top}(I_{n}+DL)\\
 & =R+\underset{\mathrm{err}}{\underbrace{D^{\top}+D^{\top}DL+R^{\top}DL}}\;.
\end{align*}
We can further bound the error according to

\begin{align*}
\|\mathrm{err}\|_{2} & \le\overset{D^{\top}+R^{\top}DL}{\overbrace{\sqrt{2q+1}\|\Delta_{\phi}\|_{2}(1+\|R\|_{2}\|L\|_{2})}}+\overset{D^{\top}DL}{\overbrace{v(2q+1)\|\Delta_{\phi}\|_{2}^{2}}}\\
 & =\sqrt{2q+1}(1+v\|\pi\|_{2})\|\Delta_{\phi}\|_{2}+v(2q+1)\|\Delta_{\phi}\|_{2}^{2}
\end{align*}
\begin{align*}
\|\mathrm{err}\|_{F}^{2} & \le n(2q+1)\|\Delta_{\phi}\|_{2}^{2}+nv(2q+1)\|\Delta_{\phi}\|_{2}^{2}+n(2q+1)\|R\|_{2}^{2}\|L\|_{2}^{2}\|\Delta_{\phi}\|_{2}^{2}\\
 & =n(2q+1)\|\Delta_{\phi}\|_{2}^{2}\big(1+v+\|\pi\|_{2}^{2}v^{2}\big)
\end{align*}
Now recall from Corollary \ref{cor:phi_bound} we have
\begin{align*}
\|\hat{\phi}-\phi\|_{\infty} & \le c_{1}q^{2}v\delta(n,p^{\tau})\\
\mathcal{E}_{\phi}:=\{\|\Delta_{\phi}\|_{2}^{2} & \le c_{1}^{2}q^{3}v^{2}\delta^{2}(n,p^{\tau})\}
\end{align*}
We thus have $\|\mathrm{err}\|_{F}^{2}\le6c_{1}^{2}\|\pi\|_{2}^{2}q^{4}v^{4}\log(p^{\tau})$,
i.e. not growing in $n$ and $\|\mathrm{err}\|_{2}\le3c_{1}\|\pi\|_{2}q^{2}v^{2}\delta(n,p^{\tau})$
for $q^{1/2}\|\Delta_{\phi}\|_{2}\le1$, the latter which occurs with
sufficient samples. It remains to put these in the deviation bound
(c.f. Lemma 1) to derive the final choice of $\bar{\lambda}$, which
leads to the final FGLS error bound. The probability below is understood
to be conditional on events used in the argument to this point, i.e.
to ensure the RE condition holds, and the AR error is bounded represented
by the event $\mathcal{E}_{\phi}$. Conditional on these we have
\begin{align*}
P[n^{-1}\|X^{\top}\hat{R}Lu\|_{\infty}\ge\bar{\lambda}/2\:|\:\mathcal{E}_{\phi}] & \le2p\exp\bigg(-c\min\bigg[\frac{\bar{\lambda}^{2}n^{2}}{4K_{X}^{2}K_{u}^{2}[\|R\|_{F}^{2}+6c_{1}^{2}\|\pi\|_{2}^{2}q^{4}v^{4}\log(p^{\tau})]},\\
 & \quad\frac{\bar{\lambda}n}{2K_{X}K_{u}[\|R\|_{2}+3c_{1}\|\pi\|_{2}q^{2}v^{2}\delta(n,p^{\tau})]}\bigg]\bigg)\;.
\end{align*}
Let $a_{n}^{2}=\|\pi\|_{2}^{2}(1+6c_{1}^{2}q^{4}v^{4}\delta^{2}(n,p^{\tau}))$
then we choose
\[
\bar{\lambda}=\frac{2K_{X}K_{u}}{c^{1/2}}\|\pi\|_{2}\big[1+3c_{1}q^{2}v^{2}\delta(n,p^{\tau})\big]\sqrt{\frac{\log(p^{\tau})}{n}}
\]
which enables us to choose the quadratic tail, and obtain $P[n^{-1}\|X^{\top}\hat{R}Lu\|_{\infty}\ge\bar{\lambda}/2\:|\:\mathcal{E}_{\Delta_{\phi}}]\le2p^{1-\tau}$,
applying the union bound (with Corollary \ref{cor:phi_bound}) gives
us 
\[
P[n^{-1}\|X^{\top}\hat{R}Lu\|_{\infty}\le\bar{\lambda}/2]\le1-20p^{1-\tau}\;.
\]
Applying this choice of $\bar{\lambda}$ in conjunction with the argument
used for Eq. \ref{eq:lasso_l2_err} gives the final result.

\end{proof}

\subsection{Proof of Lemmata for Proposition \ref{prop:re_subgaussian}}

\begin{proof}{Lemma \ref{lemma:deviation_design} - Deviation Bound (Rotated Design)} 

Expand the norm of $n^{-1/2}R\bar{W}$ in terms of inner products
as
\[
\frac{1}{n}\|R\bar{W}z\|_{2}^{2}=\frac{1}{\|\pi\|_{2}^{2}}\frac{1}{n}\sum_{t=1}^{n}\langle\sum_{s=1}^{t}R_{ts}W_{s,\cdot},z\rangle
\]
and from here down we consider the size of this random variable for
all $z\in\mathbb{S}^{p-1}$. Expanding the projection onto the sphere
we find
\begin{align*}
\langle\sum_{s=1}^{t}R_{ts}W_{s,\cdot},z\rangle^{2} & =\bigg(\sum_{j=1}^{r}\sum_{s=1}^{t}R_{ts}W_{sj}z_{j}\bigg)^{2}\\
 & =\bigg(\sum_{s=1}^{t}R_{ts}w_{s}\bigg)^{2}\;,
\end{align*}
where we let $w_{s}=\sum_{j=1}^{r}W_{sj}z_{j}$. To bound the deviations,
we first assess the sub-Gaussian norm of $w_{t}$, noting that if
$W_{sj}$ is $K_{W}$ sub-Gaussian then $w_{t}$ is sub-Gaussian with
norm $K_{w}=K_{W}\|z\|_{2}=K_{W}$. Due to the off-diagonal (overlapping
support) of rows in $R$, the terms in the above summation are dependent
which rules out direct application of the Bernstein inequality. Instead,
let us consider the vector $w=(w_{1},\ldots,w{}_{n})$, then we have
\[
\frac{1}{n\|\pi\|_{2}^{2}}\langle\sum_{s=1}^{t}R_{ts}W_{s,\cdot},z\rangle^{2}=w^{\top}R^{\top}Rw
\]
with $w$ an isotropic sub-Gaussian (zero-mean) vector with expectation
\begin{align*}
\frac{\mathbb{E}[y^{\top}R^{\top}Ry]}{n\|\pi\|_{2}^{2}} & =\frac{1}{n\|\pi\|_{2}^{2}}\mathrm{tr}(R^{\top}R)\;.
\end{align*}

Using our bounds on the trace, we have
\[
1-\frac{q}{n}\le\mathbb{E}\left[\frac{1}{n}\|R\bar{W}z\|_{2}^{2}\right]\le1\;.
\]
Applying the Hanson-Wright inequality (Lemma \ref{lemma:hanson_wright}),
we find
\begin{align*}
P[|\frac{1}{n}\|R\bar{W}z\|_{2}^{2}-1|>\epsilon+q/n] & \le2\exp\left[-c\min\left[\frac{\epsilon^{2}n^{2}\|\pi\|_{2}^{2}}{K_{W}^{4}\|R^{\top}R\|_{F}^{2}},\frac{\epsilon n\|\pi\|_{2}}{K_{W}^{2}\|R^{\top}R\|_{2}}\right]\right]\\
 & \le2\exp\left[-c\min\left[\frac{\epsilon^{2}n^{2}\|\pi\|_{2}^{2}}{K_{W}^{4}\|R^{\top}R\|_{F}^{2}},\frac{\epsilon n\|\pi\|_{2}}{K_{W}^{2}\|R^{\top}R\|_{2}}\right]\right]\\
 & \le2\exp\left[-c\min\left[\frac{\epsilon^{2}n}{K_{W}^{4}\|\pi\|_{2}^{2}(2q+1)^{2}},\frac{\epsilon n}{K_{W}^{2}\|\pi\|_{2}(2q+1)}\right]\right]\\
 & =2\exp\left\{ -c\frac{\epsilon n}{K_{W}^{2}\|\pi\|_{2}(2q+1)}\min[\frac{\epsilon}{K_{W}^{2}\|\pi\|_{2}(2q+1)},1]\right\} 
\end{align*}
Therefore if $\epsilon\le K_{W}^{2}\|\pi\|_{2}(2q+1)$ we choose the
Gaussian tail. To simplify the term involving $q/n$ we have used
$\epsilon=\epsilon'-q/n$ and found a lower bound on the tail in terms
of $\epsilon'$. Specifically 
\[
P[|\frac{1}{n}\|R\bar{W}z\|_{2}^{2}-1|>\epsilon']\le2\exp\left[-c\frac{(\epsilon'-q/n)^{2}n}{K_{W}^{4}\|\pi\|_{2}^{2}(2q+1)^{2}}\right]
\]
and we note $(\epsilon'-q/n)^{2}n\ge\epsilon'{}^{2}n/2$ under the
condition that $\epsilon'\ge4q/n$, giving
\[
P[|\frac{1}{n}\|R\bar{W}z\|_{2}^{2}-1|>\epsilon']\le2\exp\left[-c\frac{\epsilon'{}^{2}n}{2K_{W}^{4}\|\pi\|_{2}^{2}(2q+1)^{2}}\right]\;.
\]
Note, in the main statement we simplify the denominator to $K_{W}^{4}\|\pi\|_{2}^{2}(2q+1)^{2}$
to $K_{W}^{4}\|\pi\|_{2}^{2}(q+1)^{2}$ where the difference can be
put down to a change in the absolute constant $c$. For convenience
in the proofs, we will define the constant $C:=c^{-1}K_{W}^{4}\|\pi\|_{2}^{2}(q+1)^{2}$.

\end{proof}

\begin{proof}{Lemma \ref{lemma:UUP_post_rot} - Post Rotation UUP}

The strategy now is to use a covering argument based on epsilon nets
(we use notation $\eta$-net) to look at spaces spanned by various
supports, indexed by elements in $\mathcal{S}$. Given the index set
$\mathcal{S}\subset[p]$ of size $|\mathcal{S}|=r$ we consider the
space spanned by the canonical basis, i.e. $\mathcal{B}_{\mathcal{S}}=\mathrm{span}(b_{j}\:|j\in\mathcal{S})$,
now consider the space $\mathcal{U}_{\mathcal{S}}=\Sigma^{1/2}\mathcal{B}_{\mathcal{S}}$
and then intersect this with spheres to obtain $\mathcal{Z}_{\mathcal{S}}:=\mathcal{U}_{\mathcal{S}}\cap\mathbb{S}^{p-1}$,
for each subset $\mathcal{S}$ we can then consider an $\eta$-net
$\Pi_{\mathcal{S}}\subset\mathcal{Z}_{\mathcal{S}}$ where we know
that $|\Pi_{\mathcal{S}}|\le(1+2/\eta)^{r}$. Now consider $\Pi=\cup_{|\mathcal{S}|=r}\Pi_{\mathcal{S}}$
for $\eta\le1$ we have $|\Pi_{\mathcal{S}}|\le(3/\eta)^{r}$ and
${p \choose r}$ sets $\Pi_{\mathcal{S}}$ to consider, resulting
in a covering number 
\[
|\Pi|\le\exp\left(r\log(\frac{3\mathrm{e}p}{r\eta})\right)\;.
\]
Let us denote $\tilde{W}_{n}=n^{-1/2}R\bar{W}$, applying Lemma \ref{lemma:deviation_design}
we obtain

\begin{align}
P[\exists z\in\Pi\:|\:|\|\tilde{W}_{n}z\|_{2}^{2}-1|>\epsilon] & \le|\Pi|P[|\|\tilde{W}z\|_{2}^{2}-1|>\epsilon]\label{eq:deviation_z_pi}\\
 & \le2\exp\left(r\log(\frac{3\mathrm{e}p}{r\eta})\right)\exp\left(-C\epsilon^{2}n\right)\nonumber \\
 & \le2p^{-\tau}\nonumber 
\end{align}
where the last inequality holds for all 
\[
n\ge C^{-1}\left[\frac{\log(p^{\tau})+r\log(\frac{3\mathrm{e}p}{r\eta})}{\epsilon^{2}}\right]
\]
We extend from the argument from $z\in\Pi$ to all $z\in\mathcal{Z}_{\mathcal{S}}$
noting that $\|\tilde{W}_{n}z\|_{2}^{2}$ is a Lipschitz function
in $z$. Specifically, we consider 
\begin{align*}
\|\tilde{W}_{n}\|_{2,\mathcal{S}} & :=\sup_{z\in\mathcal{Z}_{\mathcal{S}}}\|\tilde{W}_{n}z\|_{2}\\
 & \le1+\epsilon/2\;,
\end{align*}
where the second line follows from (\ref{eq:deviation_z_pi}). Consider
the point(s) $z_{\Pi}\in\Pi$ closest to $z$ in the $\eta$-net.
We have
\[
\|\tilde{W}_{n}z_{\Pi}\|_{2}-\|\tilde{W}_{n}(z-z_{\Pi})\|_{2}\le\|\tilde{W}_{n}z\|_{2}\le\|\tilde{W}_{n}z_{\Pi}\|_{2}+\|\tilde{W}_{n}(z-z_{\Pi})\|_{2}
\]
note that $\|z-z_{\Pi}\|_{2}\le\eta$ and therefore $\|\tilde{W}_{n}(z-z_{\Pi})\|_{2}\le\eta\|\tilde{W}_{n}\|_{2,\mathcal{S}}$,
with (\ref{eq:deviation_z_pi}) gives
\[
1-\epsilon-\eta\|\tilde{W}_{n}\|_{2,\mathcal{S}}\le\|\tilde{W}_{n}z\|_{2}\le1+\frac{\epsilon}{2}+\eta\|\tilde{W}_{n}\|_{2,\mathcal{S}}\;,
\]
in probability greater than $1-2p^{-\tau}$. Further noting $\|\tilde{W}_{n}z\|_{2,\mathcal{S}}\le1+\frac{\epsilon}{2}+\eta\|\tilde{W}_{n}\|_{2,\mathcal{S}}$
gives $\|\tilde{W}_{n}\|_{2,\mathcal{S}}\le\frac{1+\epsilon/2}{1-\eta}$.

Assuming $q/n<\epsilon<\min(2^{-1},K_{W}^{2}\|\pi\|_{2}q)$ and choosing
$\eta=\epsilon/(1+2\epsilon)$ gives for any $\mathcal{S}$ of size
$r$
\[
1-2\epsilon<\|\tilde{W}_{n}z\|_{2}<1+2\epsilon\quad\forall z\in\mathcal{Z}_{\mathcal{S}}\;.
\]
Recalling $\|z\|_{2}=1$ for all $z\in\mathcal{Z}_{\mathcal{S}}$,
we can extend this to any sparse vector $u$ for $\mathrm{supp}(u)=\mathcal{S}$
via rescaling, i.e.
\[
\frac{Au}{\|Au\|_{2}}\in\mathcal{Z}_{\mathcal{S}}\;,
\]
giving $(1-2\epsilon)\|Au\|_{2}\le\|\tilde{W}_{n}Au\|_{2}\le(1+2\epsilon)\|Au\|_{2}$
as required.

\end{proof}

\subsection{Additional Experimental Results}

\begin{figure}[h]
\begin{centering}
\includegraphics[width=0.6\columnwidth]{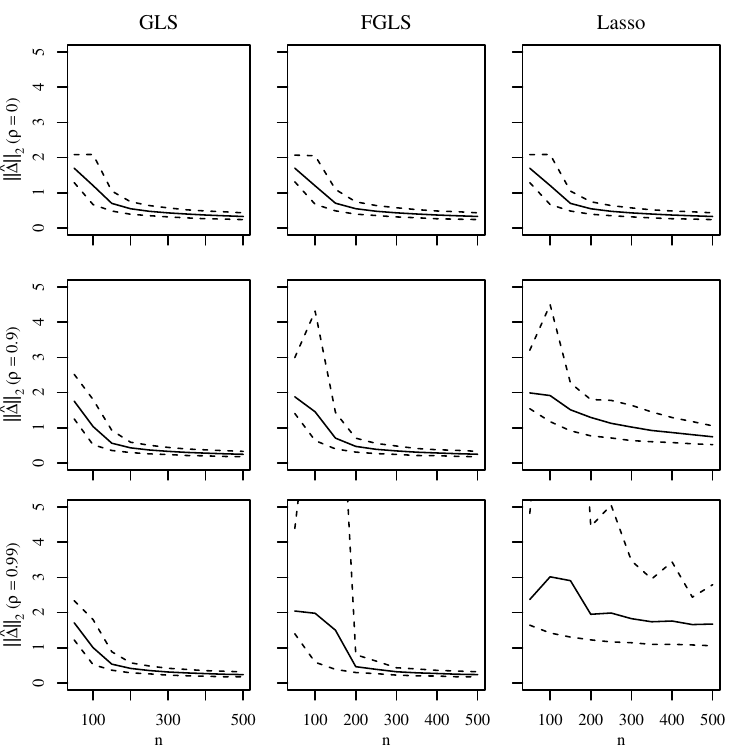}
\par\end{centering}
\begin{centering}
\includegraphics[width=0.6\columnwidth]{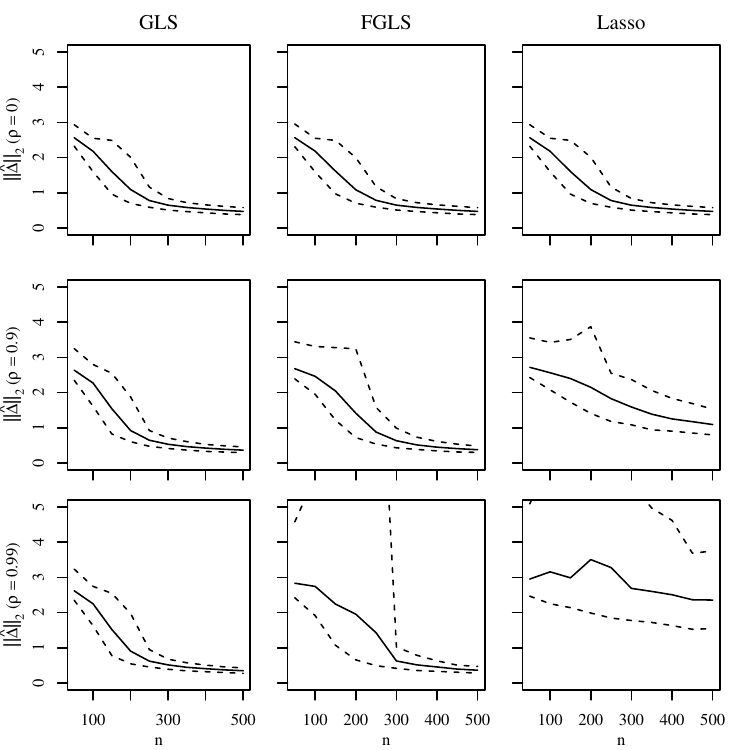}
\par\end{centering}
\caption{Estimation error $p^{-1/2}\|\hat{\Delta}\|_{2}$ as a function of
$n$ for different settings of $\rho$ (top $p=128$, bottom $p=256$),
dashed lines indicate empirical 95\% confidence intervals.}
\end{figure}

\begin{figure}[h]
\begin{centering}
\includegraphics[width=0.6\columnwidth]{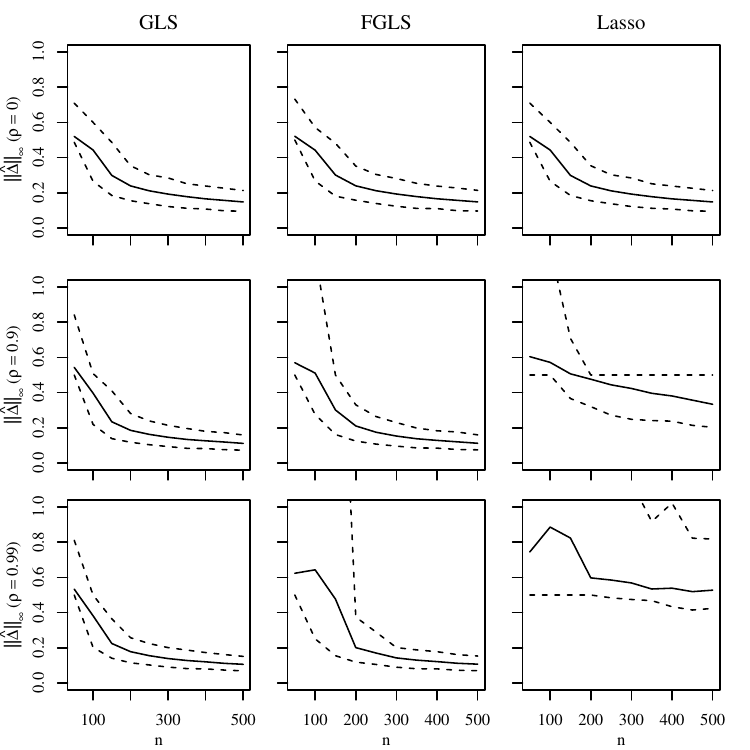}
\par\end{centering}
\begin{centering}
\includegraphics[width=0.6\columnwidth]{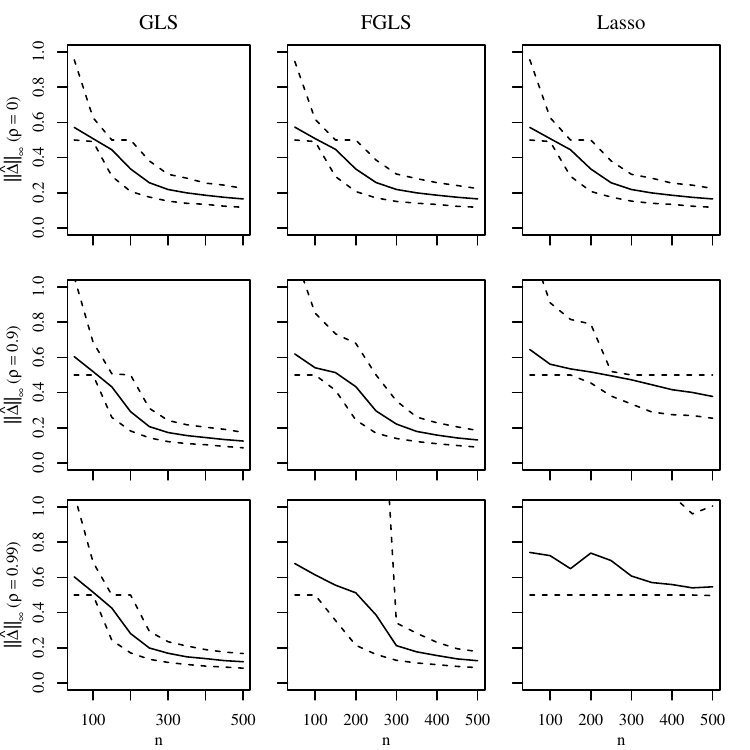}
\par\end{centering}
\caption{Estimation error $\|\hat{\Delta}\|_{\infty}$ as a function of $n$
for different settings of $\rho$ (top $p=128$, bottom $p=256$),
dashed lines indicate empirical 95\% confidence intervals.}
\end{figure}

\bibliographystyle{plain}
\bibliography{refs}

\end{document}